\documentclass[twocolumn]{aastex62}

\shorttitle{{\it NuSTAR} Observations of NGC 4968}
\shortauthors{LaMassa et al.}

\begin{document}

\title{{\it NuSTAR} Uncovers an Extremely Local Compton-thick AGN in NGC 4968}

\correspondingauthor{Stephanie LaMassa}
\email{slamassa@stsci.edu}

\author[0000-0002-5907-3330]{Stephanie M. LaMassa}
\affiliation{Space Telescope Science Institute,
3700 San Martin Drive,
Baltimore, MD 21218, USA}

\author{Tahir Yaqoob}
\affiliation{CRESST,
University of Maryland Baltimore County,
1000 Hilltop Circle, Baltimore, MD 21250, USA}
\affiliation{NASA/Goddard Spaceflight Center,
Mail Code 662,
Greenbelt, MD 20771, USA}

\author{Peter G. Boorman}
\affiliation{Department of Physics \& Astronomy,
  Faculty of Physical Sciences and Engineering,
  University of Southampton,
  Southampton SO17 1BJ, UK}
\affiliation{Astronomical Institute,
  Academy of Sciences,
  Bo\v{c}n\'{i} II 1401, CZ-14131 Prague, Czech Republic}

\author{Panayiotis Tzanavaris}
\affiliation{CRESST, 
University of Maryland Baltimore County,
1000 Hilltop Circle, Baltimore, MD 21250, USA}
\affiliation{NASA/Goddard Spaceflight Center,
Mail Code 662,
Greenbelt, MD 20771, USA}

\author{N. A. Levenson}
\affiliation{Space Telescope Science Institute,
3700 San Martin Drive,
Baltimore, MD 21218, USA}

\author{Poshak Gandhi}
\affiliation{Department of Physics \& Astronomy,
  Faculty of Physical Sciences and Engineering,
  University of Southampton,
  Southampton SO17 1BJ, UK}

\author{Andrew F. Ptak}
\affiliation{NASA/Goddard Spaceflight Center,
Mail Code 662,
Greenbelt, MD 20771, USA}

\author{Timothy M. Heckman}
\affiliation{Department of Physics and Astronomy, Center for Astrophysical Sciences, Johns Hopkins University, Baltimore, MD 21218, USA}

\begin{abstract}
  We present the analysis of {\it Chandra} and {\it NuSTAR} spectra of NGC 4968, a local (D$\sim$44 Mpc) 12$\mu$m-selected Seyfert 2 galaxy, enshrouded within Compton-thick layers of obscuring gas. We find no evidence of variability between the {\it Chandra} and {\it NuSTAR} observations (separated by 2 years), and between the two {\it NuSTAR} observations (separated by 10 months). Using self-consistent X-ray models, we rule out the scenario where the obscuring medium is nearly spherical and uniform, contradicting the results implied by the $<$10 keV {\it Chandra} spectrum. The line-of-sight column density, from intervening matter between the source and observer that intercepts the intrinsic AGN X-ray emission, is well within the Compton-thick regime, with a minimum column density of $2\times10^{24}$ cm$^{-2}$. The average global column density is high ($> 3\times10^{23}$ cm$^{-2}$), with both Compton-thick and Compton-thin solutions permitted depending on the X-ray spectral model. The spectral models provide a range of intrinsic AGN continuum parameters and implied 2-10 keV luminosities ($L_{\rm 2-10keV,intrinsic}$), where the higher end of $L_{\rm 2-10keV,intrinsic}$ is consistent with expectations from the 12$\mu$m luminosity ($L_{\rm 2-10keV,intrinisc} \sim 7\times10^{42}$ erg s$^{-1}$). Compared with Compton-thick AGN previously observed by {\it NuSTAR}, NGC 4968 is among the most intrinsically X-ray luminous. However, despite its close proximity and relatively high intrinsic X-ray luminosity, it is undetected by the 105 month {\it Swift}-BAT survey, underscoring the importance of multi-wavelength selection for obtaining the most complete census of the most hidden black holes.

\end{abstract}


\section{Introduction} 

Supermassive black holes in the centers of galaxies grow by accreting nearby matter in a phase where they are observed as active galactic nuclei (AGN). A significant fraction of AGN are obscured \citep[e.g.,][]{treister2004} by a parsec-scale dusty torus and sometimes by the host galaxy itself \citep[e.g.,][]{buchner2017, circosta}. Black hole growth hidden behind the most extreme, Compton-thick levels of obscuration (i.e., $N_{\rm H} > 1.25 \times10^{24}$ cm$^{-2}$) is an important component in the overall census of AGN demographics. The quoted fraction of Compton-thick AGN varies in the literature \citep[e.g., from $\sim9 - 35$\%;][]{gilli07, treister09, akylas12, brightman_ueda, ricci, buchner, ueda14}, with recent population synthesis models indicating that the number can be as high as 50\% at $z < 0.1$ \citep{ananna}.

Signatures of Compton-thick obscuration are present in the X-ray spectra of these AGN. The combined effects of photoelectric absorption and Compton down scattering of the intrinsic AGN power law continuum produce a broad Compton hump at $\sim$30 keV \citep[e.g.,][]{comastri}. However, even in the most Compton-thick AGN, the X-ray spectrum below 10 keV is not completely absorbed, as would be assumed by a simplistic absorbed power law model, since contributions from scattered AGN flux and the host galaxy are usually observed \citep[e.g.,][]{turner,winter,lamassa2012}. One of the most prominent features in the X-ray spectra of Compton-thick AGN is the neutral Fe K$\alpha$ narrow emission line at 6.4 keV, which forms in distant, cold matter \citep{ghisellini}, usually ascribed to the putative torus in AGN unification models \citep{antonucci, urry}. While the line itself forms via fluorescence in distant matter, the direct component of the continuum against which it is measured is depressed, causing the equivalent width (EW) of the line to reach 1 keV in Compton-thick AGN \citep{matt}, and to sometimes exceed several keV in the most extreme cases \citep{levenson}. However, a large Fe K$\alpha$ EW is not a defining characteristic of the most obscured AGN as Compton-thick AGN have been shown to exhibit a range of EW value that extend below 1 keV \citep{boorman2018}.

Proper modeling of the physical processes in AGN that self-consistently accounts for the transmitted emission, Compton-scattered reflected component, and fluorescent line flux is required to accurately determine the column density of the X-ray reprocessor and recover the intrinsic X-ray flux. Such modeling may also constrain the geometry of the obscurer and the level of homogeneity or patchiness. Within the past several years, a handful of X-ray spectral models created via Monte Carlo simulations with a range of assumed geometries (spherical to toroidal to clumpy) were released to the community \citep[e.g.,][]{mytorus,brightman,liu_li,borus} and can be used with conventional X-ray spectral fitting packages. Some of these models can accommodate a patchy obscuring medium, rather than a homogeneous distribution of matter assumed in previous X-ray spectral models, by allowing the global average column density of the absorber ($N_{\rm H,global}$) to be fit independently from the gas column that intercepts the transmitted emission along the line-of-sight ($N_{\rm H,los}$).

However, despite the utility of high-quality, broad band X-ray spectra for deriving these intrinsic AGN parameters, X-ray selection as a means to identify Compton-thick AGN candidates is hindered by severe attenuation of their observed X-ray flux. Fortunately, multi-wavelength diagnostics can recover Compton-thick AGN missed by X-ray selection as they rely on intrinsic AGN proxies that are to first order unaffected by the column density of the obscuring medium, arising from scales more spatially extended than that of the X-ray obscuring gas \citep[e.g.,][]{lamassa2010}. One such diagnostic is the mid-infrared flux arising from AGN heated dust \citep{gandhi,levenson2009,asmus}, which was used to create the {\it IRAS} 12$\mu$m-selected sample of Seyfert 2 galaxies \citep{12um_samp}.

A Compton-thick AGN candidate from this 12$\mu$m sample is NGC 4968, an extremely local ($z=0.00986$; D $\sim$ 44 Mpc) Seyfert 2 galaxy, which is remarkable for its extreme Fe K$\alpha$ EW value \citep[EW=$2.5^{+2.6}_{-1.0}$ keV;][]{lamassa2011,me_ngc4968}. A 50 ks {\it Chandra} observation of the source revealed the coexistence of extended soft emission which was well accommodated by a thermal model (suggesting it arises from circumnuclear star formation) and Compton-thick levels of obscuration based on fitting the X-ray spectrum with physically motivated models that assume a spherical absorption geometry \citep[BNSphere;][]{brightman} and a toroidal geometry with a fixed opening angle \citep[MYTorus;][]{mytorus,yaqoob2012}. \citet{me_ngc4968} posited that the X-ray obscuring gas had a high covering factor, favoring the spherical absorption model over a toroidal geometry since this model best fit the Fe K$\alpha$ emission line. Such a geometry facilitates the production of an extreme Fe K$\alpha$ EW since the differential extinction between the fluorescent line emission and continuum is enhanced in a closed geometry compared with a toroidal geometry. Furthermore, the implied intrinsic 2-10 keV luminosity based on the mid-infrared 12$\mu$m luminosity from ALLWISE \citep{mainzer} and the $L_{\rm 12\mu m}/L_{\rm 2-10 keV}$ relation \citep{gandhi,asmus} was consistent with the intrinsic luminosity calculated from the spherical absorption model, while a toroidal absorption model provided a lower intrinsic X-ray luminosity.

As the {\it Chandra} spectrum is limited to energies below 10 keV, the data precluded \citet{me_ngc4968} from favoring the spherical geometry over the toroidal model, where extrapolations of the best fit models showed clear differences above 10 keV (see Figure \ref{comp_highE} in Appendix A). Despite its close proximity, NGC 4968 had not been detected at energies above 10 keV, and is missing from the 105-month Swift-Burst Alert Telescope (BAT) survey \citep[14-195 keV flux limit = $8.4\times10^{-12}$ erg s$^{-1}$ cm$^{-2}$;][]{oh}. Studying the high energy properties of this source required dedicated follow-up observations. We were awarded two 20 ks {\it NuSTAR} observations of NGC 4968 in Cycle 3 (PI: LaMassa) to cover the high energy X-ray spectrum of NGC 4968 to constrain the geometry of the reprocessor and accurately measure the column density and intrinsic AGN continuum. The observations were separated by 10 months to allow us to search for any variability in the high-energy spectrum between epochs, as was observed in the prototypical Compton-thick Seyfert 2 galaxy NGC 1068 \citep{marinucci}.

\section{Observations}
        {\it NuSTAR} consists of two co-aligned mirrors that focus incoming hard X-rays (3-79 keV) onto two focal plane modules, FPMA and FPMB. {\it NuSTAR} has an angular spatial resolution of 18$^{\prime\prime}$ (full-width half maximum) over a 12$^{\prime} \times 12^{\prime}$ field of view \citep{harrison}.
        
        NGC 4968 was first observed by {\it NuSTAR} on 2017 June 26 for 21 ks and was observed again on 2018 April 26 for 20 ks (see Table \ref{obs_log}). The data were reduced with the {\it NuSTAR} Data Analysis Software (NuSTARDAS) v.1.8.0 {\it nupipeline} script, which creates filtered events files. The spectra were extracted from a 30$^{\prime\prime}$ radius aperture centered on the source, while the background was extracted from a circular region with a 90$^{\prime\prime}$ radius from a source free region on the detector. Source and background spectra were extracted separately for FPMA and FPMB. {\it NuSTAR} detected net (i.e., background subtracted) counts of 232.7 and 276.3 between the two modules for the observations from 2017 and 2018, respectively. The data were grouped by a minimum of 10 counts per bin for spectral fitting.
        
        In this analysis, we include the {\it Chandra} ACIS-S spectrum from 2015, which had a total of 707.9 net counts. As discussed in \citet{me_ngc4968}, the data were reduced with the CIAO \citep{fruscione} script {\textsc chandra\_repro} to produce a filtered events file, using CIAO v. 8, with CALDB v. 4.7.1. A spectrum was extracted from a 6$^{\prime\prime}$ radius circular aperture to encompass the extended soft emission, with a background spectrum extracted from an annulus centered on the AGN and having an inner and outer radius of 10$^{\prime\prime}$ and 30$^{\prime\prime}$, respectively. The spectrum was grouped by a minimum of 5 counts per bin to provide better resolution around the Fe K$\alpha$ line than we would achieve with grouping by a higher number of counts.

\begin{deluxetable*}{llcll}
\tablecaption{\label{obs_log} Summary of X-ray Observations of NGC 4968}
\tablehead{
  \colhead{Observatory} & \colhead{Date}  & \colhead{Exposure Time}   & \colhead{Net Counts\tablenotemark{1}} & \colhead{ObsID\tablenotemark{2}} \\
  & & \colhead{ks} &
  }
\startdata
{\it Chandra}\tablenotemark{3} & 2015-March-9  & 50 & 707.9 & 17126 \\
{\it NuSTAR}                   & 2017-June-26  & 21 & 232.7 & 60302006002 \\
{\it NuSTAR}                   & 2018-April-26 & 20 & 276.3 & 60302006004 
\enddata
\tablenotetext{1}{Net counts are background subtracted. The {\it NuSTAR} counts represent the sum between the FPMA and FPMB detectors.}
\tablenotetext{2}{Observation identification number.}
\tablenotetext{3}{Data first published in \citet{me_ngc4968}.}
\end{deluxetable*}        

\section{Data Analysis}
We simultaneously fit the {\it Chandra} and {\it NuSTAR} spectra of NGC 4968 in XSpec v 12.9.1p \citep{arnaud} using the Cash statistic (C-Stat) which is more appropriate in the low count regime \citep{cash}, although the statistic as implemented in XSpec is not designed to work on background-subtracted spectra; however with net counts above 100 in each spectrum, we expect this choice of fitting statistic will not heavily bias our results (i.e., see \href{https://asd.gsfc.nasa.gov/XSPECwiki/statistical\_methods\_in\_XSPEC/}{statistical\_methods\_in\_XSPEC}).

First, we use phenomenological modeling within the common energy range of {\it Chandra} and {\it NuSTAR} to search for evidence of variability between spectral epochs. We then fit the X-ray spectra of NGC 4968 with self-consistent, physically motivated models, including BNSphere and MYTorus \citep[as was done in][]{me_ngc4968} and a model that assumes a toroidal geometry with a variable opening angle \citep[borus02\_v170323a, hereafter ``borus02'';][]{borus}. Both toroidal absorption models can accommodate a clumpy obscuring medium by allowing the line-of-sight column density (from matter in the torus intercepting the intrinsic continuum) to be fit independently from the global average column density, which is responsible for Compton scattering and fluorescent line emission out of the line-of-sight. From this analysis, we measure the obscuring column density, assess the geometry of the X-ray reprocessor, and calculate the observed and intrinsic X-ray luminosities. All errors represent the 90\% confidence interval ($\Delta$C-stat = 2.7 for one interesting parameter).

\subsection{Testing for Variability Among Observations}
We simultaneously fit the {\it Chandra} and {\it NuSTAR} FPMA and FPMB spectra within a restricted energy band that is common between the two observatories (3-8 keV) using a phenomenological model of a powerlaw plus a Gaussian component to accommodate the Fe K$\alpha$ emission line. A multiplicative constant factor was allowed to be free in the fitting to act as a cross calibration normalization for the {\it NuSTAR} spectra: any significant deviations above the $\sim$10\% cross-calibration factor \citep{madsen} between {\it Chandra} and {\it NuSTAR} observations could signal variability. Similarly, a significant discrepancy between the cross-normalization factors of the {\it NuSTAR} observations between 2017 and 2018 would indicate variability at higher energies. The other fit parameters, i.e., power law slope ($\Gamma$) and normalization, Gaussian line energy, $\sigma$, and normalization, are linked among the spectral datasets during the fitting process.

The cross calibration normalization factors we find are 0.98$^{+0.37}_{-0.31}$ and 1.16$^{+0.43}_{-0.35}$ for the FPMA and FPMB spectra, respectively, for the observation from 2017, and 0.83$^{+0.32}_{-0.27}$ and 1.39$^{+0.47}_{-0.38}$ for the FPMA and FPMB spectra for the observation from 2018. Thus, there is no evidence of variability in the overall flux levels between the {\it Chandra} and {\it NuSTAR} observations, nor between the {\it NuSTAR} spectral epochs. Furthermore, we do not see systematic discrepencies in spectral shape between the observing epochs: linking the powerlaw and Gaussian model parameters across the datasets provided an adequate description of the spectra.

\subsection{Modeling the X-ray Spectra}
Since we ruled out the possibility that variability between observations can significantly impact our results, we fit all five spectra simultaneously in the modeling analysis below to best constrain the fit. In Appendix B, we describe the details of the spectral fitting and highlight the main results here.

We fit between 3 - 50 keV in the {\it NuSTAR} spectra, adopting a cut-off at 50 keV since the spectra are background dominated above these energies. For the {\it Chandra} spectrum, we use the spectral range 0.6 - 8 keV when fitting the BNSphere and MYTorus models, and 1 - 8 keV when fitting the borus02 model since this model is not tabulated below 1 keV. In all models, we include a thermal component (\textsc{apec}) and a scattered power law to properly fit the soft emission \citep[0.6-2 keV;][]{me_ngc4968}. This latter component describes AGN continuum emission that leaks through the obscuring medium and then scatters off an extended, optically-thin zone into our line-of-sight. Thus its power law index ($\Gamma$) and normalization parameters are linked to those of the intrinsic AGN power law model, with a constant factor left free to represent the scattered light fraction ($f_{\rm scatt}$). 

We report the observed 2-10 keV and 10-40 keV luminosities ($L_{\rm 2-10keV,observed}$ and $L_{\rm 10-40keV,observed}$, respectively), where the errors on the luminosity are derived from the uncertainty on the power law normalization. From the best fit power law model parameters ($\Gamma$ and power law normalization), we calculate the intrinsic 2-10 X-ray luminosity ($L_{\rm 2-10keV,intrinsic}$), and estimate the errors on $L_{\rm 2-10keV,intrinsic}$ by calculating the minimum and maximum luminosity from the upper and lower bounds on the power law parameters found by the XSpec ``steppar'' command. We emphasize that these errors only account for the statistical error on the fit and do not include systematic errors. Throughout, we adopt a cosmology where $H_{0}$ = 67.8 km s$^{-1}$ Mpc$^{-1}$, $\Omega_{M}$ = 0.31, and $\Omega_{\Lambda}$ = 0.69 \citep{planck}. 

\subsection{Assessing Geometry of the Obscurer}

We fitted the X-ray spectra of NGC 4968 with several state-of-the art models that assume different geometries for the obscuring and reprocessing gas, including models where the medium is homogeneous and has a spherical distribution \citep[BNSphere;][]{brightman} or a toroidal geometry \citep[coupled MYTorus model;][]{mytorus}. We test whether the obscuring gas is patchy, and not necessarily a smooth ``torus,'' by using two X-ray spectral models where the line-of-sight and global average column density can be fit independently (decoupled MYTorus and borus02, \citealt{borus}) and are thus allowed to have significantly different values as would be expected from a clumpy distribution of matter.

\subsubsection{Spherical Absorption Geometry}\label{sphere}
The earlier results from fitting the {\it Chandra} spectrum alone indicated that a spherical absorption geometry \citep{brightman}, with a small percentage of light that leaks through the obscuration, was a more likely description of the observed spectrum than a toroidal geometry \citep{me_ngc4968}. In light of the {\it NuSTAR} data above 10 keV, is this interpretation still valid?

\begin{figure*}
  \includegraphics[scale=0.24]{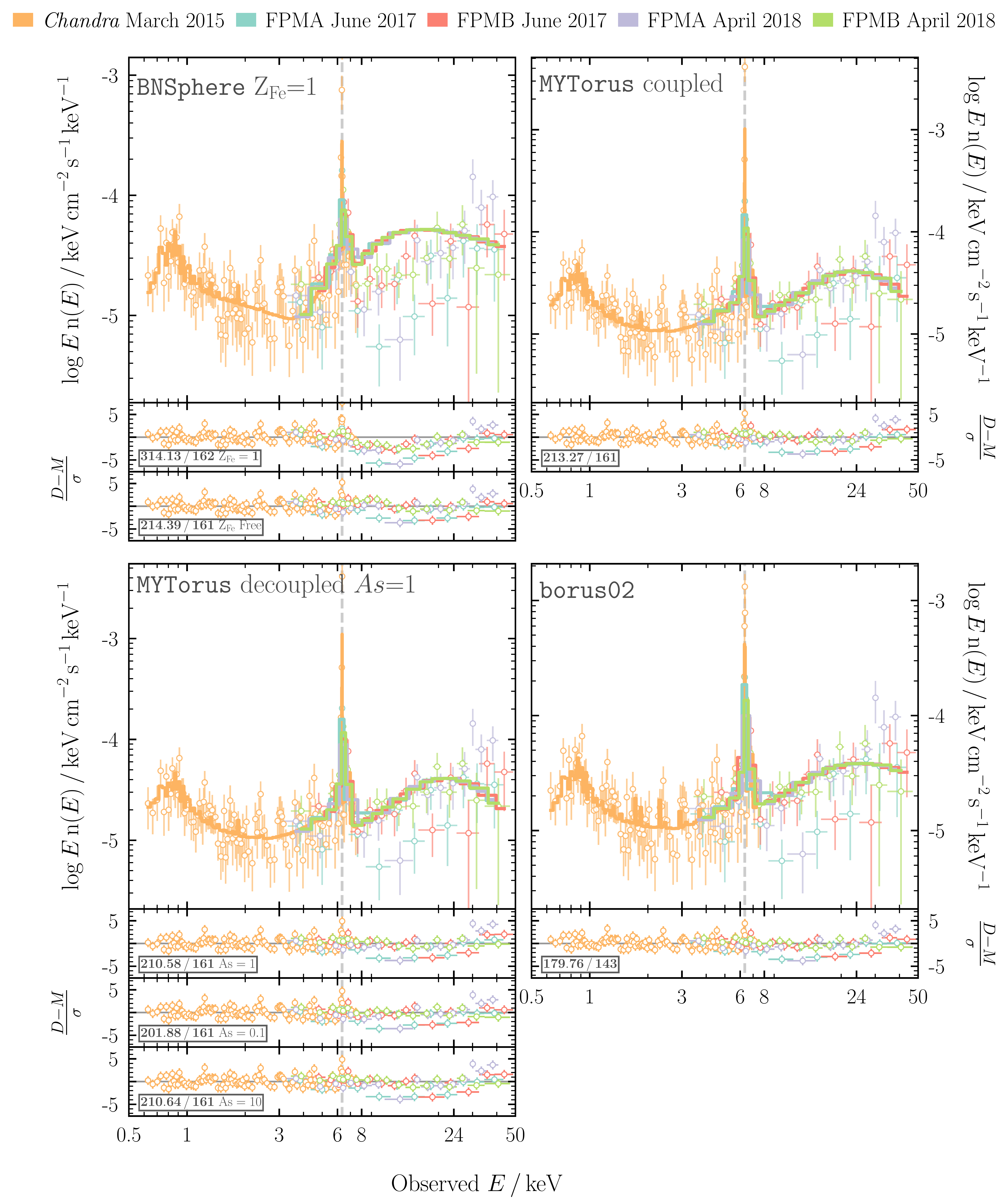}
  \caption{\label{spec_fits} Physically-motivated model fits and residuals to the {\it Chandra} and {\it NuSTAR} FPMA and FPMB spectra of NGC 4968. All models include a thermal component and scattered AGN emission to adequately model the X-ray emission between 0.5 - 2 keV. {\it Top left}: Spherical absorption model of \citet{brightman}, for both solar iron abundance and allowing the iron abundance to be a free parameter (bottom residuals panel, (D - M)/$\sigma$, where D is data, M is model, and $\sigma$ is error). With a solar Fe abundance, large residuals are present around Fe K$\alpha$, and though the fit improves when the Fe abundance is a free parameter, the abundance pegs at the maximum allowed value for the model (10 times solar), which is unphysical (see Section \ref{sphere} and Figure \ref{feka_closeup}). {\it Top right}: MYTorus spectral fit  where the model is applied in the default ``coupled'' mode (i.e., the line-of-sight and global column densities are the same, as would be expected from a homogeneous torus). The fitted column density is Compton-thick ($N_{\rm H} > 6.7\times10^{24}$ cm$^{-2}$, see Section \ref{coup_mytorus}). {\it Bottom left}:  Decoupled MYTorus spectral fits, where the line-of-sight and global column densities are fit independently for different values of $A_S$, the relative normalization between the transmitted and Compton scattered/fluorescent line emission, which is shown in the bottom(D - M)/$\sigma$  panels (see Section \ref{decoup_mytorus}). {\it Bottom right}: borus02 model fit where the global column density is fitted independently from the line-of-sight column density (see Section \ref{borus_sec}).  In all the model realizations where the line-of-sight $N_{\rm H}$ is decoupled from the global $N_{\rm H}$, we find Compton-thick column densities along the line-of-sight and average global column densities that range from heavy extinction ($N_{\rm H,global} = 4.9^{+3.0}_{-1.7} \times 10^{23}$ cm$^{-2}$ ) to being well within the Compton-thick regime (i.e., the lower limit on $N_{\rm H,global}$ is  3.2 $\times 10^{24}$ cm$^{-2}$) as summarized in Table \ref{uncoup_params}.}
  \end{figure*}

\begin{figure}
  \includegraphics[scale=0.24]{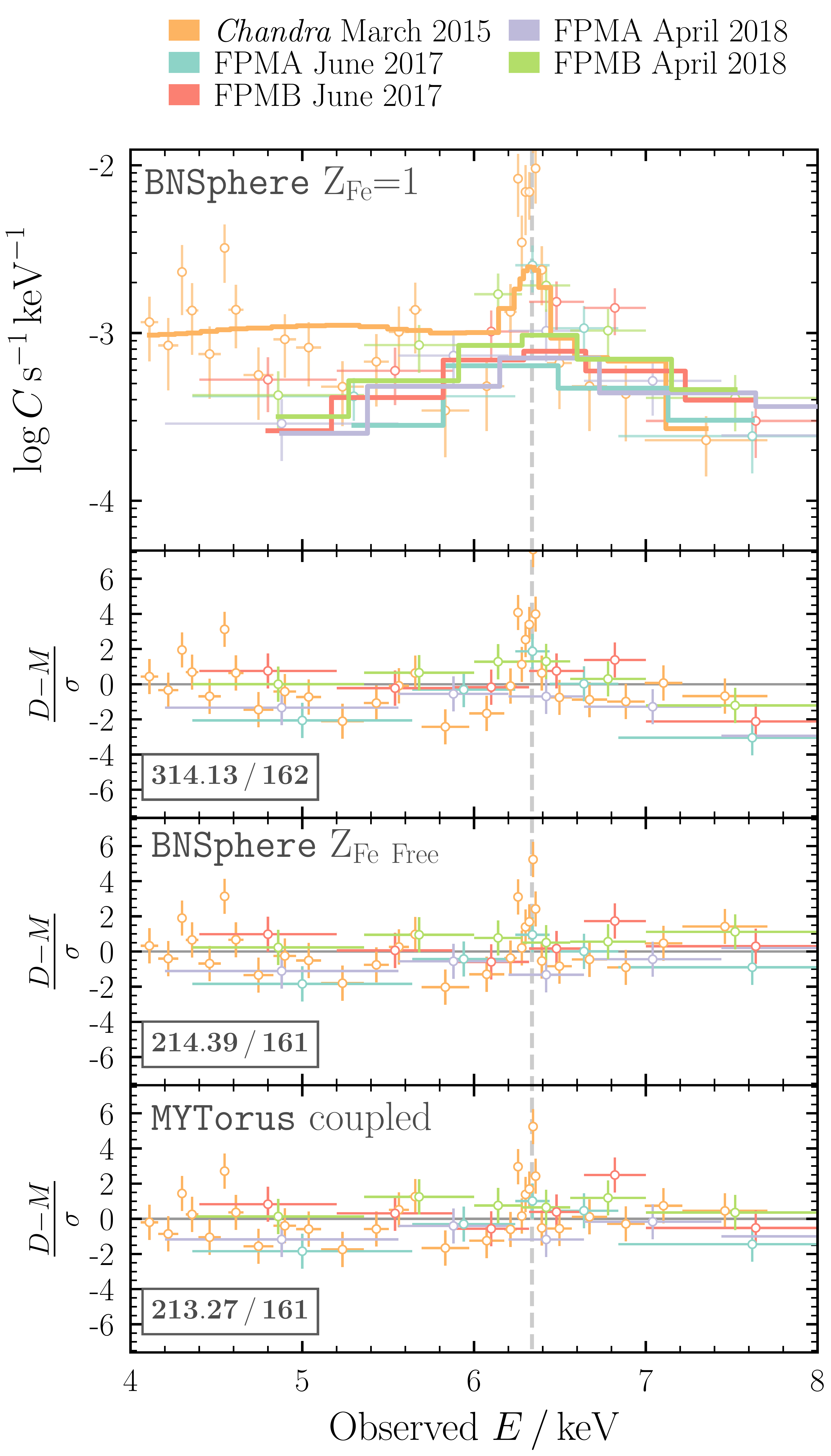}
  \caption{\label{feka_closeup} From top to bottom, close up of the spectral counts ($C$) around the Fe K$\alpha$ line when fitted with the spherical absorption model with solar Fe abundance \citep{brightman}, spherical absorption model with variable Fe abundance (which pegged at the maximum allowed value of 10 times solar), and toroidal absorption model \citep[MYTorus;][]{mytorus} with a uniform column density. The constraints from a spherical geometry prevent the continuum around the Fe K$\alpha$ line and the flux of the line from being well modeled. The toroidal model provides a better fit, both in fitting the continuum and the line, since Compton-thick column densities are permitted by this model without violating the constraints of the spectrum above 10 keV (e.g., see Figure \ref{comp_highE}).}
\end{figure}

We initially fit the spectrum with the Fe abundance frozen at solar (Figure \ref{spec_fits}, top left). The best fit paramaters are summarized in Table \ref{coup_params}. The fit is poor (C-Stat = 314 for 162 degrees of freedom), especially at the higher energies probed by {\it NuSTAR}. Large residuals around the Fe K$\alpha$ line are also apparent (see Figure \ref{feka_closeup}, top).

This model fails because in order to produce the observed Fe K$\alpha$ EW, the column density would need to be Compton-thick, which would impose a prominent Compton hump above 10 keV in a spherical geometry that is not observed in the data (see Appendix A). The model therefore struggles to produce a strong enough Fe K$\alpha$ feature when constrained by the observed high energy spectrum, resulting in a poor fit to the data. We note that in this model, the measured column density corresponds to the radial column density, $N_{\rm H,radial}$, or the amount of obscuring gas seen through the radius of a uniform sphere.

Large Fe K$\alpha$ EW values can also result from super-solar abundances, so we thawed the iron abundance to test whether we can achieve a better fit to the spectrum. The lower residual panel of Figure \ref{spec_fits} (top left) shows a better fit to the spectra, with $\Delta$C of 100 indicating a statistically superior fit due to a deeper Fe K edge, and there is a greater line flux (Figure \ref{feka_closeup}, middle). However, the fitted Fe abundance soars to the maximum allowed table value in the spherical absorption model (i.e., 10 times solar), indicating that this model realization is unphysical.

We thus conclude that a spherical, uniform obscuring medium with a small percentage of intrinsic AGN leakage, our preferred model based on the {\it Chandra} spectrum \citet{me_ngc4968}, is disfavored based on the constraints from the higher energy coverage of {\it NuSTAR}. The extreme Fe K$\alpha$ EW value requires Compton-thick levels of obscuration, which would result in a strong Compton hump in a spherical geometry that is not observed in the data, or a super-solar Fe abundance, where the model requires an unphysical solution to fit the spectra.

\begin{deluxetable*}{llll}
\tablecaption{\label{coup_params} Best-Fit Spherical Absorption and Coupled MYTorus\tablenotemark{1} Model Parameters}
\tablehead{
  \colhead{Parameter} & \colhead{Sphere}     & \colhead{Sphere}  & \colhead{MYTorus} \\
                      & \colhead{Frozen Fe}  & \colhead{Free Fe} & \colhead{(coupled)}
}
\startdata
kT (keV)                                                & 0.73$\pm0.12$      & 0.73$^{+0.11}_{-0.12}$  &  0.77$^{+0.12}_{-0.13}$ \\
{\textsc apec} normalization(10$^{-6}$)\tablenotemark{2} & 6.35$^{+2.09}_{-1.96}$ & 7.09$^{+1.97}_{-1.93}$  & 6.16$^{+2.14}_{-2.08}$  \\
$\Gamma$                                                & 1.53$\pm0.20$       & 1.38$\pm0.18$       & 2.22$^{+0.30}_{-0.22}$  \\
$N_{\rm H,radial}$ (10$^{24}$ cm$^{-2}$)                      & 0.53$^{+0.11}_{-0.08}$ & 0.24$\pm0.05$       & \nodata    \\
$N_{\rm H,equatorial}$ (10$^{24}$ cm$^{-2}$)                   & \nodata            & \nodata             & $>6.7$ \\
$\theta_{\rm incl}$\tablenotemark{3}                       & \nodata            & \nodata             & 69.6$^{+11.1}_{-6.7}$    \\
Power law normalization(10$^{-4}$)\tablenotemark{4}      & 2.40$^{+1.79}_{-1.03}$ & 1.37$^{+0.86}_{-0.55}$ & 102$^{+618}_{-53}$   \\
Fe abund                                                & 1(f)                   & $>8.2$      & 1(f) \\
$f_{\rm scatt}$ (10$^{-2}$)\tablenotemark{5}                & 7.4$^{+4.0}_{-2.6}$    & 11.7$^{+5.7}_{-3.6}$ & 0.16$^{+0.19}_{-0.02}$ \\
C-Stat (dof)                              & 314.13 (162)        & 214.39 (161)      &  213.27 (161) \\
\hline
Log($L_{\rm 2-10keV,observed}$ erg s$^{-1}$)  & 40.9$\pm0.2$ & 40.8$\pm0.2$ & 40.8$^{+0.9}_{-0.3}$ \\
Log($L_{\rm 10,40keV,observed}$ erg s$^{-1}$  & 41.6$\pm0.2$ & 41.7$\pm0.2$ & 41.6$^{+0.9}_{-0.3}$ \\
Log($L_{\rm 2-10keV,intrinsic}$ erg s$^{-1}$) & 41.5$\pm0.4$ & 41.3$^{+0.3}_{-0.4}$ & 42.6$^{+1.0}_{-0.5}$ \\ 
\enddata
\tablenotetext{1}{The column density measured in the spherical model represents the radial column density while that measured by the MYTorus model is the equatorial column density. Integrating the incident radiation over a spherical distribution of matter gives a $N_{\rm H, radial} \sim \pi/4 \times N_{\rm H,equatorial}$ correspondence between the two parameters \citep[see][]{mytorus}. For fitted parameters from the ``decoupled'' MYTorus modeling, where the line-of-sight and global column densities are fit independently, see Table \ref{uncoup_params}. }
\tablenotetext{2}{The {\textsc APEC} normalization is given as $\frac{10^{-14}}{4\pi [D_A(1+z) ]^2} \int n_e n_H dV$, where $D_A$ is the angular diameter distance to the source in cm and $n_e$ and $n_H$ are the electron and hydrogen densities, respectively, in cm$^{-3}$.}
\tablenotetext{3}{$\theta_{\rm incl}$ represents the inclination of the X-ray obscuring torus with respect to the line-of-sight. 90$^{\circ}$ is edge-on, 0$^{\circ}$ is face-on, and values greater than $>$60$^{\circ}$ in the MYTorus model intersect the X-ray obscuring medium. }
\tablenotetext{4}{The power law normalization has units of photons keV$^{-1}$ cm$^{-2}$ s$^{-1}$ at 1 keV.}
\tablenotetext{5}{$f_{\rm scatt}$ refers to the fraction of the intrinsic AGN continuum that leaks through holes in a patchy obscuring medium and scatters off an extended optically-thin zone into our line-of-sight.\\}
\end{deluxetable*}

\subsubsection{Uniform Toroidal Obscuration: ``Coupled'' MYTorus Model}\label{coup_mytorus}
We fit the {\it Chandra} and {\it NuSTAR} spectra of NGC 4968 with the uniform (``coupled'') MYTorus model \citep{mytorus}, which in its default configuration, assumes an azimuthally symmetric torus with a circular cross-section (i.e., ``doughnut''-shaped) and a fixed opening angle of 60$^{\circ}$, corresponding to a covering factor ($C_{\rm tor}$) of 0.5.  The fitted column density represents the equatorial column density of the torus ($N_{\rm H,equatorial}$). Compared to the radial column density measured by the spherical absorption model, $N_{\rm H, equatorial} = 4/\pi \times N_{\rm H, radial}$ \citep[see][]{mytorus}. Lines of sight with inclination angles ($\theta_{\rm inc}$) greater than 60$^{\circ}$ intersect the torus while lower angles correspond to a face-on geometry.

This spectral fit is shown in Figure \ref{spec_fits} (top right) and the fit parameters are summarized in Table \ref{coup_params}. We find Compton-thick levels of obscuration, $N_{\rm H, equatorial} > 6.7\times10^{24}$  cm$^{-2}$, where the upper limit on the column density reaches the maximum allowed value in MYTorus (10$^{25}$ cm$^{-2}$).

The coupled MYTorus fit is a significant improvement over the spherical absorption model with solar Fe abundance (C-Stat = 213.27 for 161 degrees of freedom), since a toroidal geometry can accommodate the weaker Compton hump above 10 keV when the column density is Compton-thick compared with the spherical absoprtion model (see Figure \ref{comp_highE}). We also point out that the continuum around the Fe K$\alpha$ line and the line itself is better modeled with the toroidal absorption geometry when compared with the spherical absorption model (Figure \ref{feka_closeup}). Thus, it is apparent that the obscuring medium does not have a closed geometry as had been implied by just the $<10$ keV {\it Chandra} spectrum. 

\subsubsection{Modeling a Non-Uniform Obscuring Medium with the ``Decoupled'' MYTorus Model}\label{decoup_mytorus}
In the ``decoupled'' realization of the MYTorus model, the geometry is no longer constrained to be toroidal, and instead mimics a patchy, non-uniform obscuring medium. We note that in the coupled implementation of the MYTorus model where we measure the equatorial column density, the line-of-sight column density measured in the MYTorus model ($N_{\rm H,los,MYTorus}$) depends on $N_{\rm H,equatorial}$ and the inclination angle of the torus \citep[see Figure 1 and Equation 1 of][]{mytorus}; the two quantities are equivalent when the inclination angle of the torus is 90$^{\circ}$, or completely edge-on.

During the spectral fitting, when we allowed the relative normalization between the transmitted continuum and the Compton scattered and fluorescent line emission components ($A_S$) to be a free parameter, the fit was unconstrained.  We therefore set $A_S$ to a range of reasonable values (0.1, 1, 10) to explore the effects on measured column densities. We list the best fits to these model realizations in Table \ref{uncoup_params} and plot the fits in Figure \ref{spec_fits} (bottom left).

In all cases, the line-of-sight column density is Compton thick, while the best-fit global average column density is Compton-thick for $A_{S}$ = 1 and $A_{S}$=10, and Compton-thin for $A_{S}$ = 0.1. The solutions where the line-of-sight and global column densities are consistent indicate that the X-ray reprocessor around the AGN in NGC 4968 may be homeogeneous. We also see over an order-of-magnitude spread in the scattering fraction ($f_{\rm scatt}$) and power law normalization, the latter of which causes a corresponding spread in the implied intrinsic 2 - 10 keV luminosity.

\begin{deluxetable*}{lllll}
  \tablecaption{\label{uncoup_params} Best-Fit Decoupled MYTorus \& borus02 Model Parameters for Independent Global and Line-of-Sight Column Densities}
\tablehead{
  \colhead{Parameter} &  \colhead{MYTorus} & \colhead{MYTorus} & \colhead{MYTorus}           & \colhead{borus02}  \\
  & \multicolumn{3}{c}{(Decoupled)} \\
  & \multicolumn{3}{c}{\rule{5.5cm}{.02cm}} \\
                      &  \colhead{$A_S$ = 0.1} & \colhead{$A_S$ = 1} & \colhead{$A_S$ = 10}  & 
}
\startdata
kT (keV)                                            & 0.75$^{+0.11}_{-0.12}$  & 0.78$^{+0.13}_{-0.14}$  & 0.78$^{+0.13}_{-0.14}$ & 0.78 (f) \\
{\textsc apec} normalization(10$^{-6}$)              & 6.89$^{+2.26}_{-2.24}$  & 5.60$^{+2.17}_{-2.11}$   & 5.59$^{+2.12}_{-2.04}$ & 5.6 (f) \\
$\Gamma$                                            & 1.73$^{+0.31}_{-0.30}$  & 2.39$^{+0.03}_{-0.17}$  & 2.38$^{+0.03}_{-0.17}$ & 1.71$^{+0.03}_{-0.09}$ \\
$N_{\rm H,los,mytorus}$ (10$^{24}$ cm$^{-2}$)\tablenotemark{1}  & 4.79$^{+0.76}_{-0.57}$ & $>3.8$            & $>2.09$            & ... \\
$N_{\rm H,los,borus}$ (10$^{24}$ cm$^{-2}$)\tablenotemark{1}  & ... & ...           & ...           & $>3.4$ \\
$N_{\rm H,global}$ (10$^{24}$ cm$^{-2}$)\tablenotemark{1}  & 0.49$^{+0.30}_{-0.17}$  & $>7.2$                 & $>8.62$            & $>3.2$ \\
$C_{\rm tor}$\tablenotemark{2}                        & 0.5(f)              & 0.5(f)              & 0.5(f)              & 0.23$^{+0.03}_{-0.02}$ \\
Power law normalization(10$^{-4}$)                   & 130$^{+119}_{-63}$ & 98$^{+46}_{-31}$    & 9.68$^{+4.29}_{-2.99}$ & 11.2$^{+9.8}_{-0.6}$ \\
$f_{\rm scatt}$ (10$^{-2}$)                            & 0.13$^{+0.09}_{-0.05}$ & 0.18$\pm0.08$          & 1.79$\pm0.57$       & 1.43$^{+0.23}_{-0.73}$\\
C-Stat (dof)                                         & 201.88 (161)         & 210.58 (161)           & 210.64 (161)        & 179.76 (143) \\ 
\hline
Log($L_{\rm 2-10keV,observed}$ erg s$^{-1}$)  & 40.8$\pm0.3$ & 40.8$\pm0.2$  & 40.8$\pm0.2$ & 40.8$^{+0.2}_{-0.3}$ \\
Log($L_{\rm 10-40keV,observed}$ erg s$^{-1}$  & 41.6$\pm0.3$ & 41.6$\pm$0.2  & 41.6$\pm0.2$ & 41.6$^{+0.2}_{-0.3}$ \\
Log($L_{\rm 2-10keV,intrinsic}$ erg s$^{-1}$) & 43.1$\pm0.5$ & 42.5$^{+0.3}_{-0.2}$ & 41.5$^{+0.3}_{-0.2}$ & 42.0$^{+0.3}_{-0.4}$ 
\enddata
\tablenotetext{1}{$N_{\rm H,los,mytorus}$ ($N_{\rm H,los,borus}$) refers to the line-of-sight obscuration from the MYTorus (borus02) model that intercepts the direct continuum. $N_{\rm H,global}$ refers to the average global column density of the X-ray reprocessing gas.}
\tablenotetext{2}{$C_{\rm tor}$ is the covering factor of the torus, and is equal to cos($\theta_{\rm tor}$), where $\theta_{\rm tor}$ is the opening angle of the torus. $C_{\rm tor}$ is fixed at 0.5 in the MYTorus model. }
\end{deluxetable*}

\subsubsection{Toroidal Obscuration Geometry with Variable Opening Angle in a Non-Uniform Medium: borus02 Model}\label{borus_sec}
The borus02 X-ray spectral fitting model is similar in geometry to the BNTorus model of \citet{brightman}: the torus consists of a homogeneous sphere with two polar cutouts that are conical in shape, unlike the MYTorus geometry which is ``doughnut''-shaped. Unlike the MYTorus model, the covering factor of the torus is a free parameter in borus02. This model also corrects calculation errors in the original BNTorus model on which borus02 is based that were pointed out by \citet{liu}. We note that this error affects the toroidal model geometry of \citet{brightman}, where X-ray photons that reflected off the far side of the torus were not subsequently reabsorbed by the torus for intersecting sight lines. This bug does not affect the spherical absorption model BNSphere. Since the borus02 model is only valid for energies above 1 keV, we restrict the {\it Chandra} spectrum to $>$1 keV in the fitting below.

Similar to the decoupled MYTorus model, borus02 permits the line-of-sight column density ($N_{\rm H,los,borus}$) to be disentangled from the global average column density, mimicking a patchy medium. Like the decoupled MYTorus model fits, the line-of-sight column density is Compton-thick, and we find that the global average column density is consistent with the line-of-sight absorption, implying that the torus has a uniform column density. $L_{\rm 2-10keV,intrinsic}$ from the borus02 model is within the range estimated by the MYTorus models.

We plot the best fit borus02 model to the {\it Chandra} and {\it NuSTAR} spectra in Figure \ref{spec_fits} (bottom right).The fit parameters are summarized in the final column of Table \ref{uncoup_params}.

\section{Discussion}

\subsection{Comparison of X-ray Model Fits}
The main result of our simultaneous physically motivated modeling of the {\it Chandra} and {\it NuSTAR} spectra of NGC 4968 is that the X-ray reprocessor does not have a closed geometry or a high covering factor: the Compton-thick levels of obscuration needed to produce the extreme Fe K$\alpha$ EW are incompatible with the observed weak Compton hump within a nearly uniform, spherical absorption framework.

The MYTorus and borus02 model fits to the spectra (Tables \ref{coup_params} and \ref{uncoup_params}) exhibit a range of scattering fractions ($0.08\% <  f_{\rm scatt} < 2.36$\%), power law indices ($1.43 < \Gamma < 2.52$) and normalizations ($7\times10^{-4} <$ normalization $<7.2 \times 10^{-2}$), and implied intrinsic X-ray luminosities ($41.3 <$ Log($L_{\rm 2-10keV, intrinsic}$ erg s$^{-1}$) $< 43.6$). The MYTorus model solution where $A_{S}$ = 10 has the lowest powerlaw normalization ($1.0^{+0.4}_{-0.3} \times 10^{-3}$) and steepest spectral slope ($2.38^{+0.03}_{-0.17}$), resulting in the lowest $L_{\rm 2-10keV, intrinsic}$ value among the non-spherical models. We note that the data show effective excess residual emission above $\sim$25 keV compared with the model fits, which could be due to fluorescent emission lines between 22 - 32 keV from the internal background \citep{wik} that are imperfectly removed during background subtraction.

While the quality of the data and limitations from the modeling preclude us from drawing strong conclusions about the intrinsic AGN continuum, with over two orders of magnitude spread in the intrinsic X-ray luminosity, all these models which assume either a non-spherical or patchy distribution of matter agree on one point: NGC 4986 is enshrouded behind a line-of-sight column density exceeding $2\times 10^{24}$ cm$^{-2}$, well within the Compton-thick regime.

\subsection{Reflection or Transmission Dominated Emission? Clues from Variability, or Lack Thereof}
Most of the non-spherical absorption models ascribe the X-ray emission above 2 keV to reprocessed torus emission, with a negligible contribution from the direct AGN continuum (see Appendix B.2 - B.4 ). However, the decoupled MYTorus modeling with $A_{S}$ = 0.1 indicates that the X-ray spectrum above 20 keV is transmission dominated. If this scenario is true, we might expect to observe variable X-ray emission at these energies if the X-ray continuum varies, while such signatures would be diluted if the X-ray spectrum is reflection dominated.

For instance, {\it NuSTAR} observations of Seyfert 2 galaxy NGC 4945 revealed X-ray flux and spectral variability above 10 keV, with a factor of 2 variability on 20 ks timescales and factor 4 variability among three {\it NuSTAR} observations, separated by 4 months and 1 month. The spectra and light curves were well described by a model where the transmitted continuum pierces through intervening Compton-thick material and represents the primary observed emission above 10 keV \citep{puccetti2014}; the spectrum softens (i.e., $\Gamma$ becomes steeper) as the flux increases. Monitoring campaigns of Seyfert 1 galaxy NGC 4593 revealed a factor of two flux variability between 10 - 50 keV within a two-day window that is also associated with ``softer when brighter'' spectral variability \citep{ursini}. {\it NuSTAR} observations of a sample of Seyfert 1 galaxies indicate that they can vary on the timescale of hours in the 3 - 79 keV band \citep{rani}, though the relatively low number of counts above 10 keV preclude us from searching for intra-observation variability in NGC 4968.

In the 10 month window between {\it NuSTAR} observations of NGC 4968, we observed no X-ray flux or spectral variability at any energies, which favors an interpretation where the spectrum is reflection dominated above 10 keV, which is consistent with all the non-spherical model realizations except for the decoupled MYTorus model where $A_S$ = 0.1. We note, however, that spectral and flux variability above 10 keV may not always be ascribed to changes of the direct continuum observed in transmission dominated spectra: as \citet{marinucci} demonstrated, the temporary brightening of NGC 1068 above 20 keV could be explained by the transit of an eclipsing cloud with column density $N_{\rm H} \geq 2.5 \times10^{24}$ cm$^{-2}$ out of the line-of-sight, briefly unveiling the primary AGN emission.

\subsection{X-ray Constraints on Line-of-sight and Global Column Densities}
\begin{figure*}
  \includegraphics[scale=0.6]{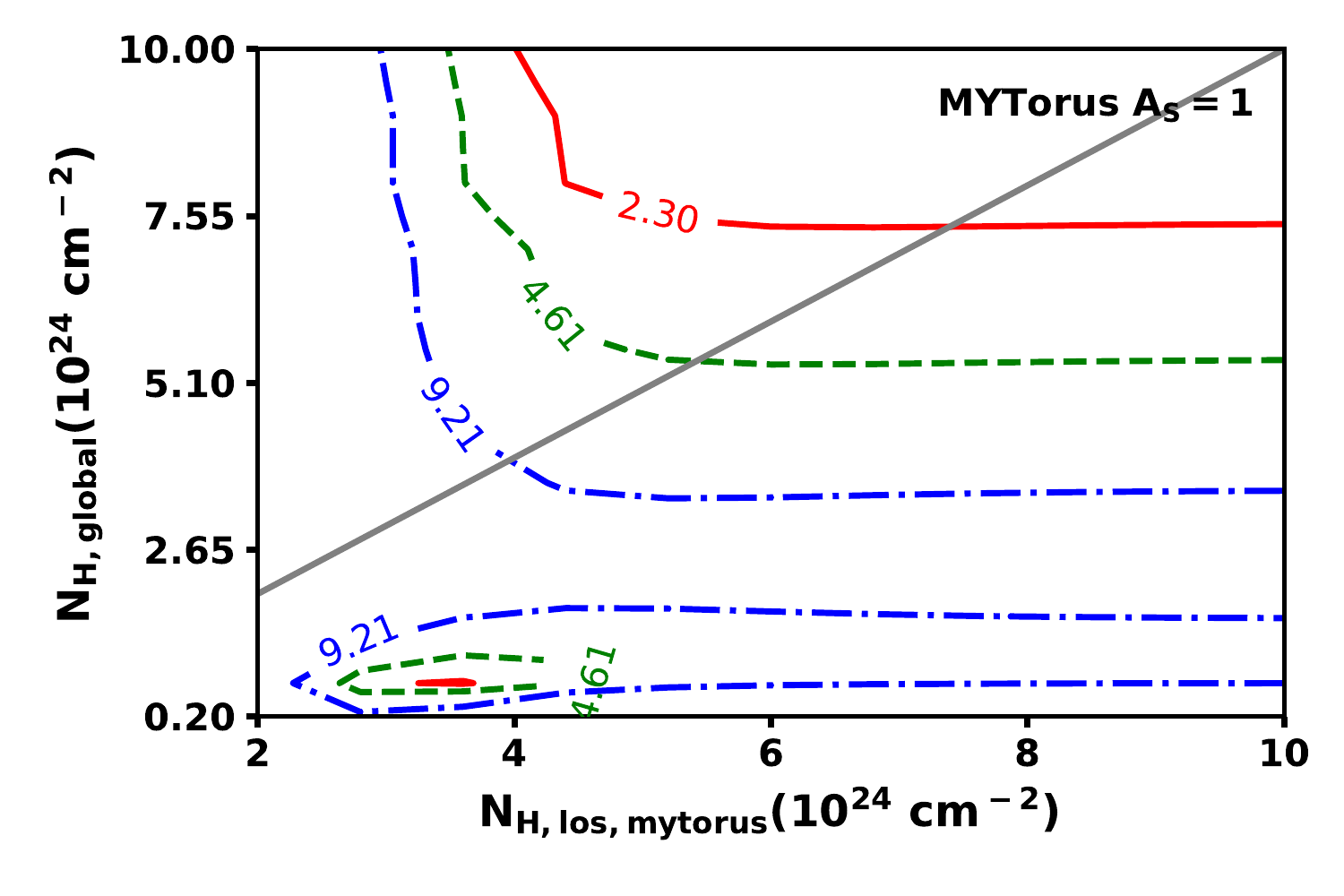}~
  \includegraphics[scale=0.6]{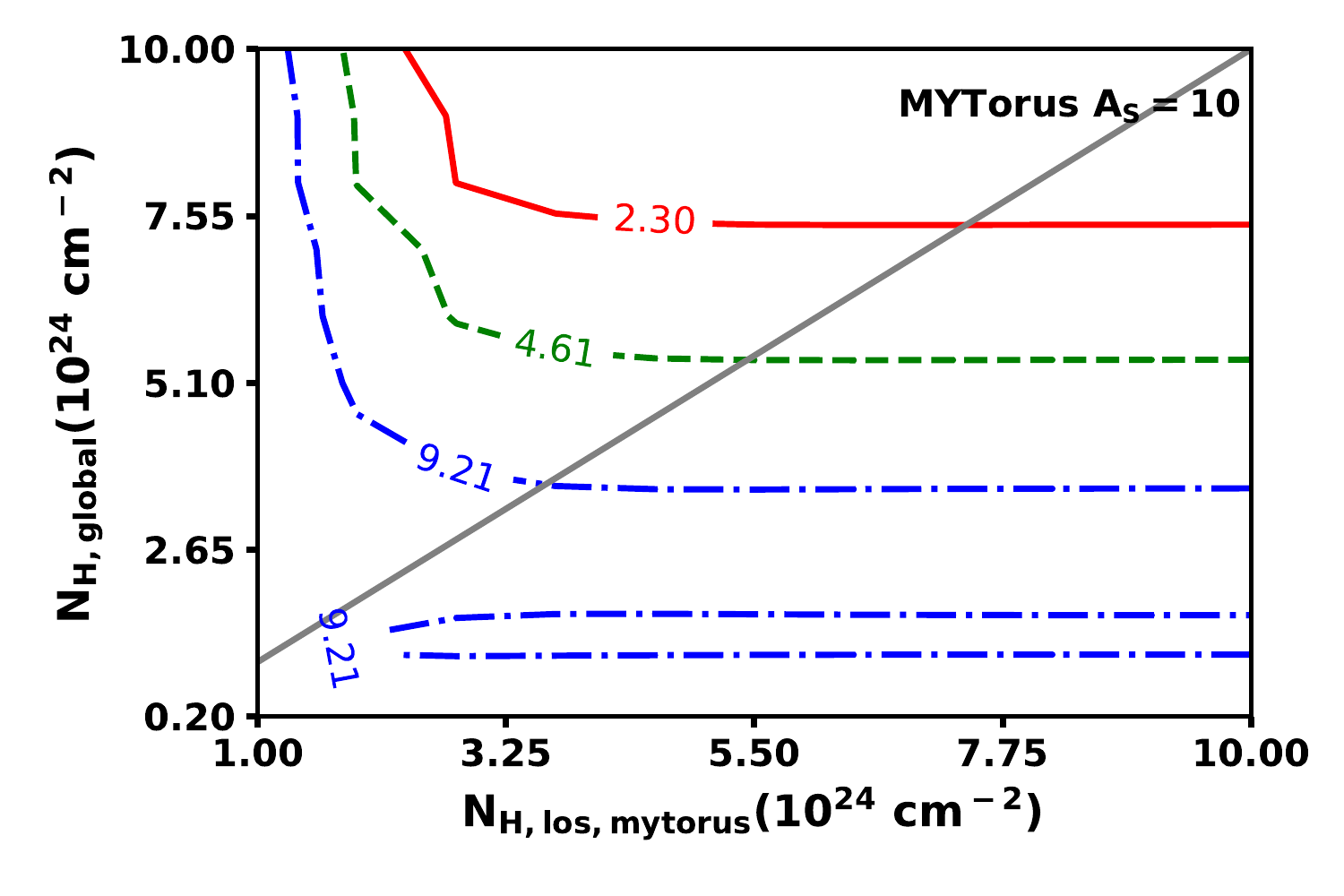} \\
  \includegraphics[scale=0.6]{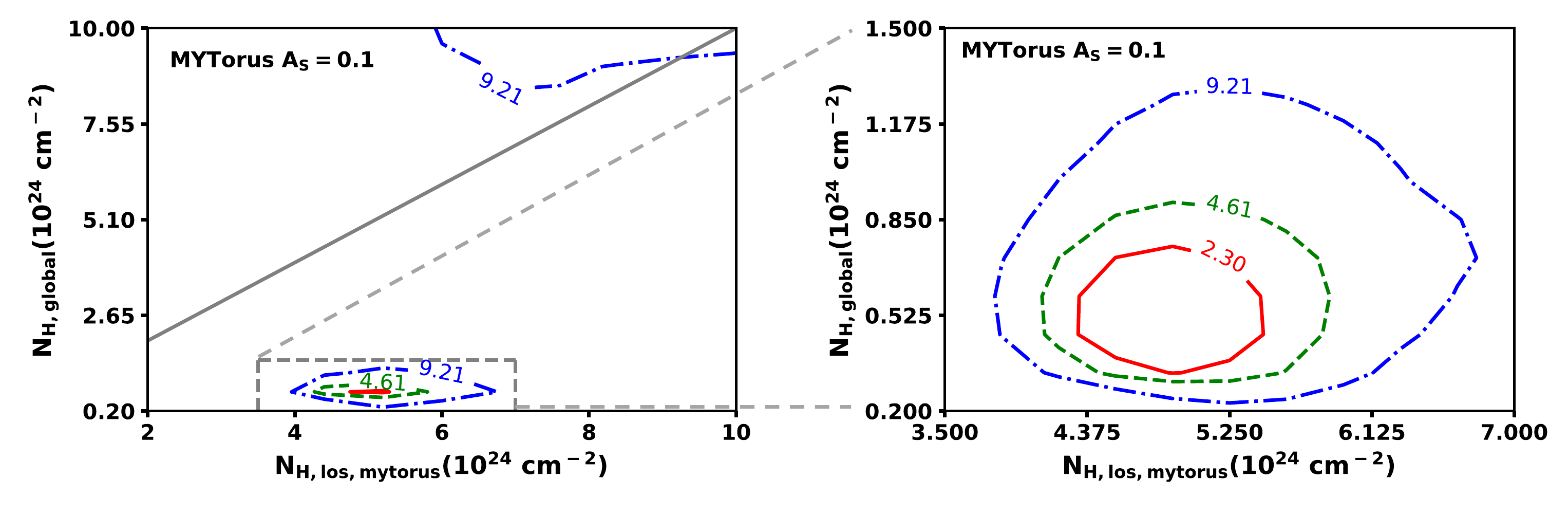}\\
  \includegraphics[scale=0.6]{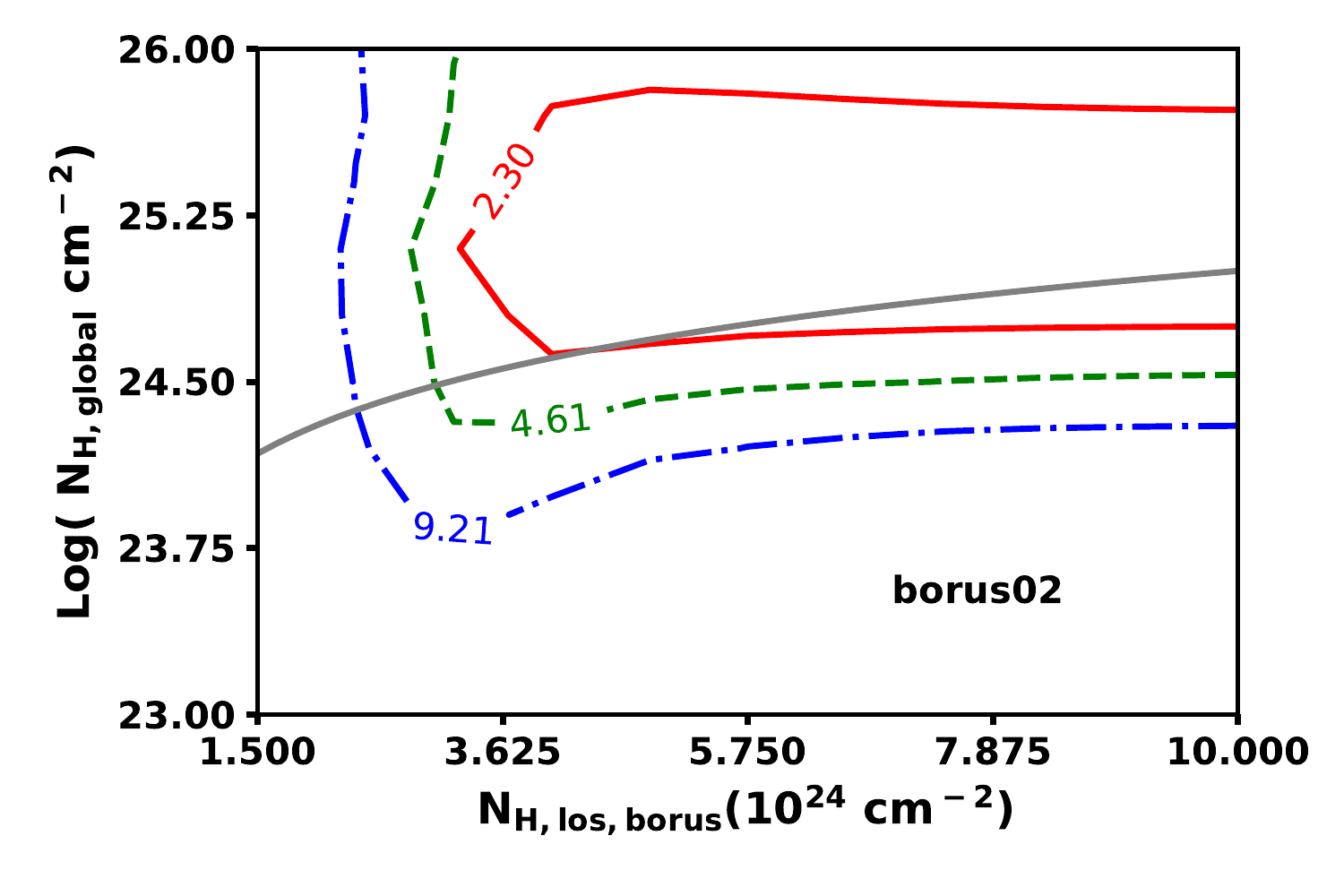}
  \caption{\label{conf_contours} Confidence contours of line-of-sight column density from the  MYTorus ($N_{\rm H,los,mytorus}$) or borus02 ($N_{\rm H,los,borus}$) models versus global column density ($N_{\rm global}$) for model realizations where the global and line of sight column densities are fitted independently. The contours correspond to $\Delta$C-Stat levels of 2.3 (68\% confidence level, red solid line), 4.61 (90\% confidence level, green dashed line), and 9.21 (99\% confidence level, blue dot-dashed line). The grey solid line denotes where the line-of-sight and global average column densities are equal. {\it Top}: MYTorus model where the relative normalization between the transmitted and Compton-scattered component ($A_{\rm S}$) is frozen to 1 ({\it left}) and 10 ({\it right}). {\it Middle}: MYTorus model where $A_{\rm S}$ is fixed to 0.1, showing the full range of values that cannot be ruled out at the 99\% confidence level ({\it left}), and a zoom-in on the best-fit values ({\it right}). {\it Bottom}: borus02 model that is set up to have the same geometry as the MYTorus model in decoupled mode. All models find that the line-of-sight column density is Compton-thick at the 99\% confidence level. Both Compton-thick and heavily obscured Compton-thin $N_{\rm H}$ values are permitted, depending on the input model assumptions. Permissible solutions include a scenario where the global and line-of-sight column densities are consistent, indicating that the the obscuring medium may be the prototypial torus with a uniform, homogeneous gas distribution or a clumpy distribution with uniform clump sizes and column densities.}
\end{figure*}

In Figure \ref{conf_contours}, we plot the confidence contours of the line-of-sight versus global column density for our assumed model geometries. Regardless of the model, the line-of-sight column density is constrained to be Compton-thick at the 99\% confidence level.

However, the best-fit values of the global column density vary based on assumptions about the normalization between the transmitted and Compton scattered emission in the MYTorus model. Though we cannot distinguish between a Compton-thin and Compton-thick solution for the global column density in the MYTorus models, all models agree that the average global column density exceeds 10$^{23}$, and extreme levels of Compton-thick obscuration ($>10^{25}$ cm$^{-2}$) cannot be ruled out.

Furthermore, we are unable to determine with certainty that the obscuring medium is nonhomogeneous as we obtain global column density values that are consistent with that along the line-of-sight in both the MYTorus and borus02 models. Results like this pose an interesting question for the definition of ``Compton-thick'' AGN, especially when simple models are used which do not have the capability to distinguish gas columns along the line-of-sight from the global average. While the transmitted AGN continuum is attenuated by Compton-thick levels of obscuring gas in NGC 4968, the global average gas density can be much lower, and indeed, lower $N_H$ values are measured by more simplistic X-ray models for NGC 4968 \citep{lamassa2011}. Models of the cosmic X-ray background that do not account for this spectral complexity likely under-predict the percentage of the most obscured black hole growth.

\subsection{Comparison to Infrared Properties}
\citet{lira} model the nuclear infrared spectral energy distribution of NGC 4968 \citep{videla} with the clumpy torus models of \citet{nenkova2008a,nenkova2008b}. They determine the infrared obscuring medium to have an inclination angle between 75$^{\circ}$ and 90$^{\circ}$ at the 67\% confidence level, with a torus half-opening angle between 52$^{\circ}$ and 68$^{\circ}$, at the same statistical confidence. The IR-derived torus half-opening angle is different than what we measure with the borus02 model ($C_{\rm tor} < $0.32 or $>$71$^{\circ}$), though as we note in Appendix B.4, other factors can influence the parameter defined as the covering factor in the borus02 model.

As discussed in \citet{me_ngc4968}, the MIR-derived column density is $\sim 2 \times10^{23}$ cm$^{-2}$, which is over an order of magnitude lower than what we measure along the {\it line-of-sight} from the X-ray spectra. The {\it global} column density measured from the X-ray spectra does permit solutions that are consistent with that from the MIR (i.e., several $\times 10^{23}$ cm$^{-2}$; Figure \ref{conf_contours}). Such an agreement may be expected as the MIR spectral modeling accounts for the ensemble of dust clouds reprocessing accretion disk photons, and not just the absorption along the line-of-sight. We point out, however, that dust-free obscuring gas within the sublimation zone of the AGN can obscure X-rays, while only more distant dusty material affects the IR emission. Thus, it is possible that the IR and X-ray reprocessor geometries are different, with a larger X-ray column density.

From the MIR 12$\mu$m to intrinsic 2-10 keV X-ray luminosity relation for local Seyfert galaxies established in \citet{asmus}:
\begin{eqnarray}\label{eqn_x_v_mir}
  \begin{array}{ll}
    {\rm Log}\left(\frac{L_{12\mu m}}{10^{43} \rm{erg}\ \rm{s}^{-1}}\right) = (0.33 \pm 0.04)\ + \\
    (0.97 \pm 0.03)\ {\rm Log}\left(\frac{L_{\rm 2-10keV,intrinsic}}{10^{43} \rm{erg}\ \rm{s}^{-1}}\right),
  \end{array}
  \end{eqnarray}
we can estimate the intrinsic 2-10 keV X-ray luminosity. Though this relationship is derived for nuclear MIR emission (using subarcsecond apertures on 8-meter class telescopes), and {\it WISE} has an angular resolution of 6.5$^{\prime\prime}$ at 12$\mu$m \citep{wright}, {\it Spitzer} spectroscopic observations indicate that the MIR emission is dominated by the central AGN and not the extended host galaxy. \citet{lamassa2010} found that the EWs of the polycyclic aromatic hydrocarbons (PAHs) at 11.3$\mu$m and 17$\mu$m were relatively weak, $\sim$0.17 and $\sim$0.1 $\mu$m, respectively, compared to sources where the MIR is dominated by star formation (EW $>$ 1$\mu$m). While the PAHs are energized by star formation processes, the boost in the MIR continuum from AGN-heated dust dilutes the PAH EW, causing weaker PAH EW values in sources where the MIR emission is dominated by the AGN.

Based on the ALLWISE \citep{wright, mainzer} $W3$ (12$\mu$m) magnitude of NGC 4968 ($5.105\pm0.014$, Vega, measured via profile-fitting photometry), $L_{\rm 12\mu m} = 1.62\pm0.05 \times 10^{43}$ erg s$^{-1}$, which implies $L_{\rm 2-10keV,intrinsic} = 7.52^{+0.80}_{-0.72} \times 10^{42}$ erg s$^{-1}$. This implied intrinsic 2-10 keV luminosity value is consistent with what we obtain with the MYTorus coupled model fit (Table \ref{coup_params}) and decoupled MYTorus model where $A_{S}$ = 0.1 and 1 (Table \ref{uncoup_params}).

We can also use the [OIV] 25.89$\mu$m line as a proxy of the intrinsic AGN luminosity \citep{melendez,rigby,diamond-stanic}, which forms in the extended narrow line region, is primarily ionized by the AGN, and is less affected by extinction than optical emission lines. For a sample of Seyfert 1 galaxies, the mean ratio of the 2-10 keV X-ray flux to [OIV] flux is $\langle$Log$\frac{F_{\rm 2-10keV}}{F_{\rm [OIV]}} \rangle = 1.92 \pm 0.6$ dex \citep{diamond-stanic}. The measured [OIV] flux for NGC 4968 from {\it Spitzer} is 2.63 $\times 10^{-13}$ erg cm$^{-2}$ s$^{-1}$ \citep{lamassa2010}, giving an expected unabsorbed 2-10 X-ray luminosity of 5.06$^{+15.08}_{-3.79} \times 10^{42}$ erg s$^{-1}$, similar to what we obtain from the 12$\mu$m luminosity estimate albeit with a larger spread.

\subsection{NGC 4968 Compared to Compton-thick AGN Studied by NuSTAR}
We place NGC 4968 in context with other ``bona-fide'' Compton-thick AGN observed with {\it NuSTAR} \citep[see][for a summary]{boorman2016} by comparing the 2-10 keV X-ray luminosity (observed and intrinsic) with the 12$\mu$m luminosity \citep[see also][]{gandhi2015}. We define ``bona-fide'' Compton-thick AGN to be those sources that lack strong X-ray variability and have Compton-thick levels of obscuration, either globally or along the line-of-sight, determined by X-ray spectral fitting with physically motivated models that includes $>$10 keV {\it NuSTAR} coverage. In Table \ref{bfct}, we list X-ray and MIR luminosities of these AGN.

\begin{deluxetable*}{lllllll}
  \tablecaption{\label{bfct} X-ray and 12$\mu$m Luminosities of Compton-Thick AGN Observed with {\it NuSTAR}\tablenotemark{1}}
\tablehead{
  \colhead{Source} &  \colhead{$L_{\rm 2-10keV,observed}$} & \colhead{Reference} & \colhead{$L_{\rm 2-10keV,intrinsic}$}  & \colhead{Reference} & \colhead{$L_{\rm 12\mu m}$} & \colhead{Reference} }
\startdata
Arp 299B     & 41.4  & \citet{ptak2015}      & 43.2  & \citet{ptak2015}      & 44.22 & {\it WISE} \\
Circinus     & 40.44 & \citet{asmus}     & 42.57 & \citet{arevalo}   & 42.65 & \citet{asmus} \\
ESO 116-G018 & 41.4  & \citet{zhao2019}      & 43.23 & \citet{zhao2019}      & 43.58 & {\it WISE} \\
IC 2560      & 40.9  & \citet{balokovic2014} & 42.9  & \citet{balokovic2014} & 43.01 & {\it WISE} \\
IC 3639      & 40.79 & \citet{boorman2016}   & 43.4  & \citet{boorman2016}   & 43.52 & \citet{asmus} \\
Mrk 34       & 42.08 & \citet{gandhi2014}   & 43.95 & \citet{gandhi2014}    & 44.2 & {\it WISE} \\
NGC 1320     & 40.95 & \citet{balokovic2014}& 42.95 & \cite{balokovic2014} & 43.15 & {\it WISE} \\
NGC 1448\tablenotemark{2}     & 38.95 & \citet{annuar2017}   & 40.74 & \citet{annuar2017}    & 42.05 & \citet{asmus} \\
NGC 2273     & 40.93 & \citet{masini2016}    & 43.11 & \citet{masini2016}    & 42.95 & \citet{asmus} \\
NGC 4945\tablenotemark{3}     & 39.85 & \citet{asmus}     & 42.74 & \citet{puccetti2014}  & 39.95 & \citet{asmus} \\
NGC 5347     & 40.53 & \citet{kammoun}   & 42.16 & \citet{kammoun}   & 43.08 & \citet{asmus} \\
NGC 5643\tablenotemark{2}     & 40.59 & \citet{asmus}     & 42.1  & \citet{annuar2015}    & 42.53 & \citet{asmus} \\
NGC 6240S    & 42.33 & \citet{asmus}     & 43.72 & \citet{puccetti2016}  & 43.56 & \citet{asmus} \\
NGC 7674\tablenotemark{2}     & 42.18 & \citet{gandhi2017}    & 43.85  & \citet{gandhi2017}    & 44.26 & \citet{asmus} \\
\enddata
\tablenotetext{1}{Luminosities are reported in log space and are in units of erg s$^{-1}$. Observed and intrinsic X-ray luminosities are from the same work, unless the observed X-ray luminosities were not reported in the studies that estimated the intrinsic X-ray luminosity. In these cases, the observed X-ray luminosities are from \citet{asmus} as listed in the table. We note that the observed X-ray luminosity is largely model independent, unlike the intrinsic X-ray luminosity.}
\tablenotetext{2}{A range of $L_{\rm 2-10keV,intrinsic}$ is reported in \citet{annuar2015,annuar2017,gandhi2017}, depending on the spectral model used to fit the data. The luminosity we list here corresponds to the mean luminosity from the models.}
\tablenotetext{3}{NGC 4945 was observed in four different spectral states by {\it NuSTAR} as described in \citet{puccetti2014}. $L_{\rm 2-10keV,intrinsic}$ listed here corresponds to the ``super-high'' state.}
\end{deluxetable*}

\begin{figure}
  \includegraphics[scale=0.58]{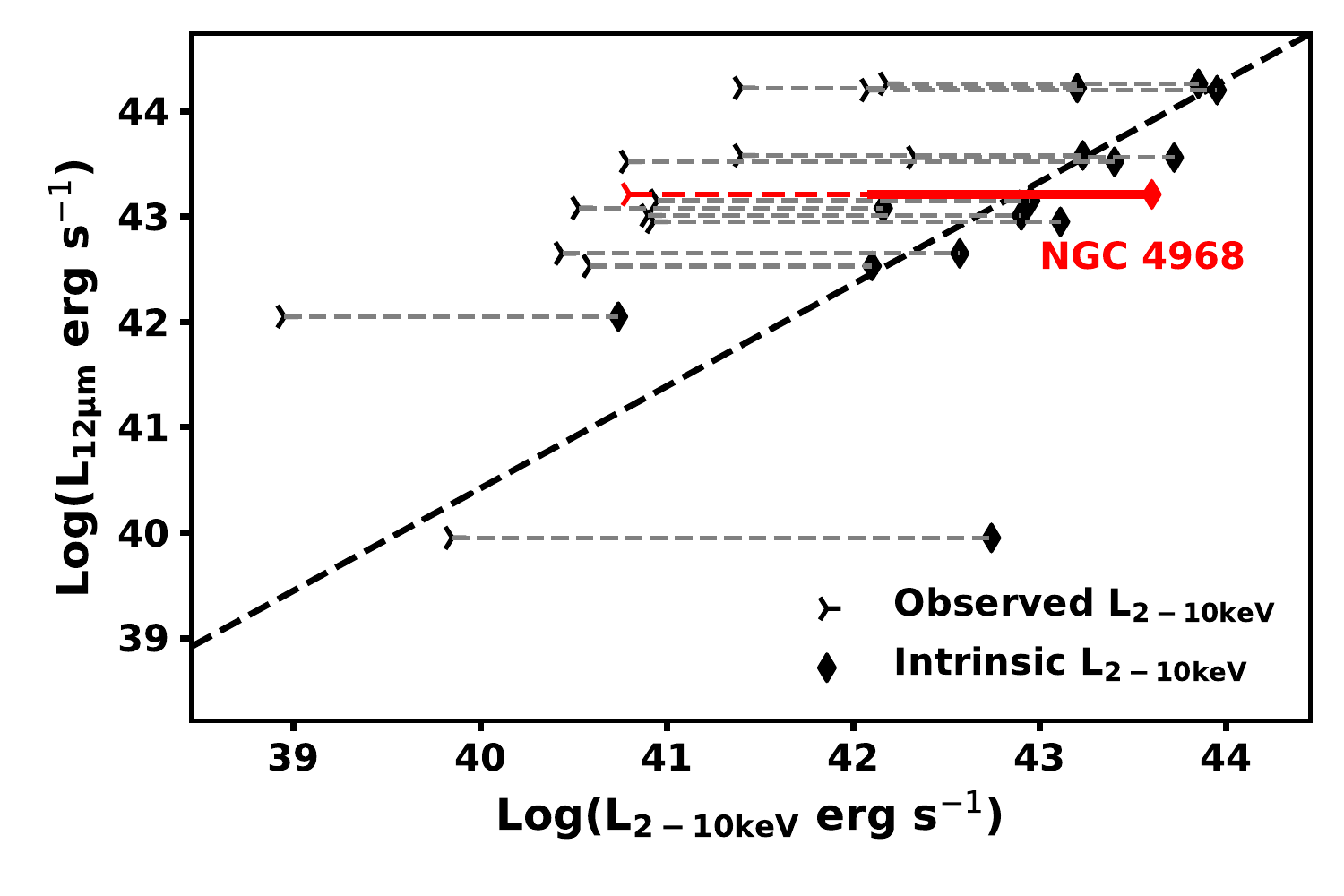}
  \caption{\label{x_v_mir} 2-10 X-ray luminosity versus 12$\mu$m luminosity for bona-fide Compton-thick AGN observed previously with {\it NuSTAR}, where the right facing caret indicates the observed luminosity, the horizontal line denotes the intrinsic luminosity and both values are connected by a grey dashed line. NGC 4968 is plotted in red, where the solid red line shows the spread in the intrinsic luminosity values from the coupled MYTorus model, with the dashed line extending to the observed 2-10 keV luminosity. The $L_{12\mu m}$/$L_{\rm 2-10keV,intrinsic}$ relation from \citet{asmus} is shown by the dashed black line for reference. The upper range on the intrinsic X-ray luminosity for NGC 4968 is consistent with expectations from the 12$\mu$m luminosity for the source. NGC 4968 is among the most intrinsically X-ray luminous Compton-thick AGN in the 2-10 keV range yet observed by {\it NuSTAR}.}
\end{figure}

In Figure \ref{x_v_mir}, we plot the range between the observed and intrinsic 2-10 keV luminosities for the archival Compton-thick AGN and the $L_{12\mu m}$/$L_{\rm 2-10keV,intrinsic}$ relation from \citet[][equation \ref{eqn_x_v_mir}]{asmus}. The spread in  $L_{\rm 2-10keV,intrinsic}$ values for NGC 4968 based on the coupled MYTorus model, which has the widest spread in the intrinsic luminosity among the values calculated from the non-spherical models, is shown by the red line, with the observed 2-10 keV luminosity noted by the red caret. As mentioned above, the upper range of $L_{\rm 2-10keV,intrinsic}$ is consistent with expectations from the observed 12$\mu$m luminosity.

In general, there is an excellent agreement between the inferred intrinsic X-ray luminosity and the 12$\mu$m luminosity with the \citet{asmus} relation for most of the Compton-thick AGN observed by NuSTAR, indicating that for this population, there is a common origin that powers both the MIR and intrinsic X-ray emission (i.e., reprocessing of accretion disk photons) to give rise to this correlation. We also highlight that NGC 4968 is among the most intrinsically X-ray luminous Compton-thick AGN yet observed with {\it NuSTAR}.

\subsection{Whither NGC 4968 in Swift-BAT?}
Despite the relatively high intrinsic X-ray luminosity and close proximity of NGC 4968, it is undetected by the all sky 105-month Swift-BAT survey, which has a 14-195 keV flux limit of $8.4\times10^{-12}$ erg s$^{-1}$ cm$^{-2}$ \citep{oh}. At the distance of NGC 4968 (D $\sim$ 44 Mpc), the fraction of Compton-thick AGN identified by Swift-BAT is consistent with the predicted ``bias-corrected'' intrinsic Compton-thick AGN fraction within the uncertainties \citep{ricci}, making the non-detection of NGC 4968 surprising.
  However, this result can be understood by the extreme levels of obscuration making the observed flux at even the highest X-ray energies heavily attenuated ($f_{\rm 14-195keV} = 3\times10^{-12}$ erg s$^{-1}$ cm$^{-2}$), pushing it below the flux limit of the BAT survey. Results like these indicate that our census of the most obscured black hole growth is underpredicted even in the nearby Universe. The topic of missing Compton-thick AGN via hard X-ray selection, correcting for this bias with mid-IR AGN selection, and the implications for the local Compton-thick AGN population will be further explored by the {\it NuSTAR} Local AGN $N_{\rm H}$ Distribution Survey (NuLANDS; Boorman et al. in prep.).

\section{Conclusions}
  We presented a joint analysis of {\it NuSTAR} and {\it Chandra} observations of nearby Seyfert 2 galaxy NGC 4968. We find no evidence of X-ray variability in the 2 year window between the {\it Chandra} observation (2015 March) and first {\it NuSTAR} observation (2017 June), nor between the two {\it NuSTAR} observations, separated by 10 months (2017 June to 2018 April).

  We used self-consistent, physically motivated X-ray models \citep[BNSphere, MYTorus, borus02;][]{brightman,mytorus,borus} to jointly fit the {\it NuSTAR} and {\it Chandra} spectra to assess the geometry of the X-ray reprocessor, gas column density, and intrinsic X-ray luminosity (Figure \ref{spec_fits}). In contrast to the results implied by fitting just the {\it Chandra} spectrum \citep{me_ngc4968}, the {\it NuSTAR} data demonstrates that a nearly spherical obscuring medium is ruled out.  In order to have a high enough column density to produce the observed Fe K$\alpha$ line, a strong Compton hump above 10 keV would be prominent in a nearly uniform spherical distribution of matter compared with a toroidal geometry due to the increased probability of multiple Compton scatterings in the former. This effect is not observed in the data. Thus the spherical absorption model of \citet{brightman} provides a poor fit to the broad band X-ray spectra, highlighting the importance of spectral coverage above 10 keV for an accurate characterization of the obscuration geometry in Compton-thick AGN.

  Models that assume a  uniform toroidal or patchy obscuring geometry provide a better fit to the X-ray spectra, however we are unable to significantly favor one model realization over another. We fit multiple configurations of the MYTorus model \citep{mytorus}, including the default ``coupled'' mode where the column density is assumed to be uniform, and the ``decoupled'' mode where the line-of-sight and average global column densities are fit independently. We also fit the X-ray spectra with the borus02 toroidal model, which accommodates a non-homeogenous obscuring medium by separately measuring the global and line-of-sight column densities.

  All models demonstrate that the line-of-sight obscuration exceeds $2 \times 10^{24}$ cm$^{-2}$, and is thus unequivocally Compton-thick. A range of global column densities are permitted by the models, with both Compton-thin and Compton-thick solutions, including solutions where the global average column density is consistent with that along the line-of-sight (Figure \ref{conf_contours}). Thus we are unable to determine whether the obscuring medium enshrouding NGC 4968 is non-uniform. We point out that even with Compton-thin solutions in the global column density, the obscuration levels are still quite high, exceeding $3\times10^{23}$ cm$^{-2}$. Our results also stress that in instances where the line of sight column density is significantly different from the global average, the definition of an AGN as ``Compton-thick'' may be ambiguous. In particular, simplistic models that do not account for the complexity of patchy obscuration are misleading, which has implications for the Compton-thick fraction assumed from fitting the cosmic X-ray background with such models.

  The models predict a range of parameters for the intrinsic AGN continuum (parameterized as a power law); scattering fraction; and implied intrinsic 2-10 keV luminosity ($L_{\rm 2-10keV,intrinsic}$), which spans a range of 2 orders of magnitude (41.3 $<$ Log ($L_{\rm 2-10keV,instrinsic}$ erg s$^{-1}$ cm$^{-2}$) $<$ 43.6, including statistical errors; see Tables \ref{coup_params} and \ref{uncoup_params}). The high range of $L_{\rm 2-10 keV, intrinsic}$ is consistent with the value implied by the $L_{12\mu m}$/$L_{\rm 2-10keV,intrinsic}$ relation \citep[$\sim7\times10^{42}$ erg s$^{-1}$;][]{asmus}, given the ALLWISE $W3$ (12$\mu$m) luminosity of 1.62$\pm0.05 \times10^{43}$ erg s$^{-1}$. Compared with other Compton-thick AGN previously observed by {\it NuSTAR}, NGC 4968 is among the most intrinsically luminous in the 2-10 keV range.

   Due to obscuration that ranges from being heavy ($> 10^{23}$ cm$^{-2}$) to Compton thick, the observed X-ray emission from NGC 4968 at all energies is heavily attenuated. As a result, it is undetected by the 105-month Swift-BAT survey \citep{oh}, despite being relatively nearby at D$\sim$44 Mpc. Fortunately, this source is identified as an AGN thanks to its optical line diagnostics (classifying it as a Seyfert 2 galaxy), and its bright mid-infrared emission, causing it to be included in the IRAS 12$\mu$m Sy2 catalog. Hence, multi-wavelength diagnostics remain of utmost importance for achieving a comprehensive census of the most obscured black hole growth. Results from the NuLANDS survey will be used to correct for this bias against the Compton-thick population in hard X-ray surveys by leveraging the power of mid-IR AGN selection (Boorman et al. in prep.).
  
\acknowledgments
We thank the referee for a thorough reading of the manuscript and constructive comments that helped us streamline the presentation. S. M. L. acknowledges support from NASA grant 80NSSC17K0631. P.~B. acknowledges financial support from the STFC and the Czech Science Foundation project No. 19-05599Y.

\vspace{5mm}
\facilities{NuSTAR, CXO}

\software{XSpec (v12.9.1p; Arnaud 1996), CIAO (v8; Fruscione et al. 2006), NuSTARDAS (v1.8.0)}

\appendix

\section{Strength of Compton Hump for Spherical and Toroidal Absorption Geometries}
\begin{figure*}
  \includegraphics[scale=0.55]{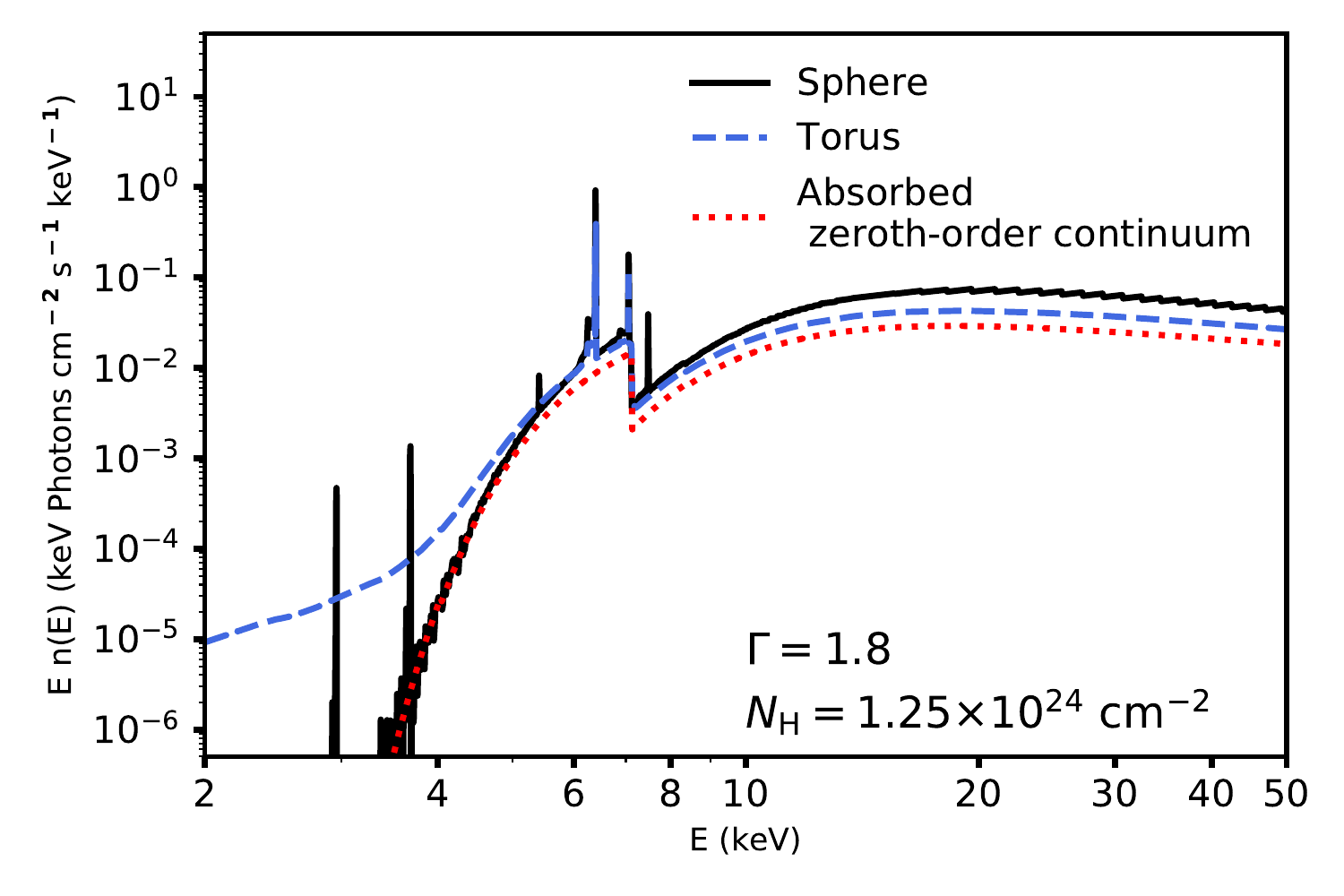}~
  \hspace{0.5cm}
  \includegraphics[scale=0.55]{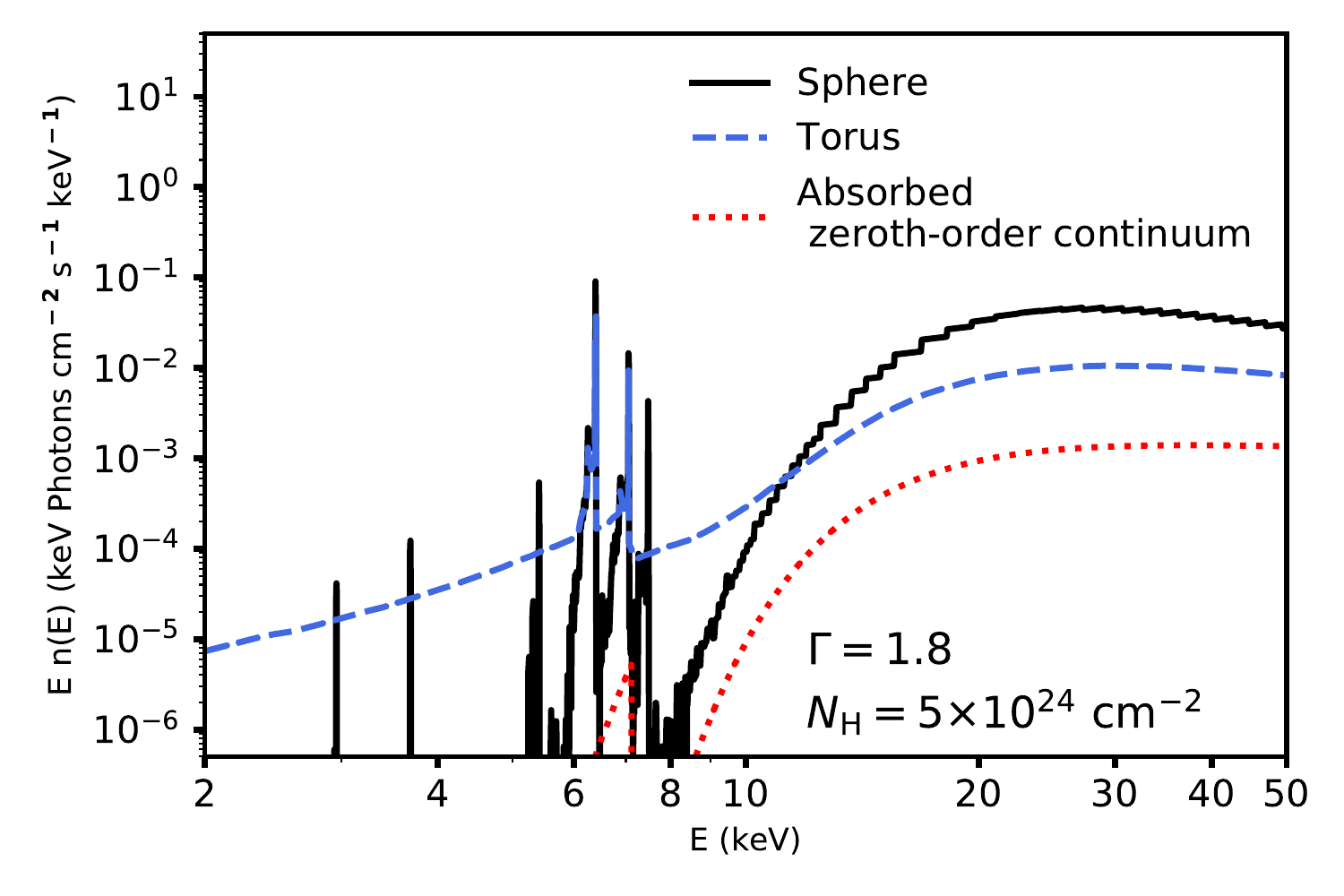}\\
  \caption{\label{comp_highE} Comparison of spherical (black solid line) and toroidal (dashed blue line) absorption model geometries with the same input AGN power law parameters ($\Gamma = 1.8$ and arbitrary normalization set to unity) for marginal Compton-thick obscuration ($N_{\rm H} = 1.25\times10^{24}$ cm$^{-2}$; left) and heavier Compton-thick levels of obscuration ($N_{\rm H} = 5\times10^{24}$ cm$^{-2}$; right). To facilitate comparison, no other model components beyond the table models from \citet{brightman} and \citet{mytorus}, which account for the absorbed transmitted continuum, Compton scattered component, and fluorescent line emission, are included. For reference, the spectrum of the absorbed, zeroth-order continuum is shown as the red dotted line (which is identical in both model geometries) to illustrate the additional effects Compton scattering has on the observed spectrum. As the column density becomes more Compton thick, the reflection spectrum above 10 keV (the energy range in which {\it NuSTAR} is sensitive) is more pronounced in a spherical geometry than in a toroidal one. This strong curvature is not observed in the {\it NuSTAR} spectrum of NGC 4968, though the Fe K$\alpha$ EW demonstrates Compton-thick levels of obscuration are present. Thus, a nearly spherical, uniform distribution of matter is an incorrect description of the X-ray reprocessor. }
\end{figure*}

 To assess the effects that Compton-thick levels of obscuration have on an AGN X-ray spectrum above 10 keV, we compare the observed spectra of a toy model with the same intrinsic power law parameters ($\Gamma = 1.8$ and arbitrary normalization of unity) within a spherical and toroidal absorber (Figure \ref{comp_highE}). For this exercise, we use the table models from \citet{brightman} and \citet{mytorus} for the spherical and toroidal model, respectively, where we assume an inclination angle of 90$^{\circ}$ for the MYTorus model. The attenuation of the direct, zeroth-order continuum is the same in both models (red dotted line), so differences in the spectra can be ascribed to the effects of Compton scattering and fluoresence within the different geometries.

We see that the spectral curvature between 10-20 keV imparted by Compton reflection is stronger in the spherical geometry compared with a toroidal one, and this discrepancy becomes more pronounced as the column density increases from marginally Compton-thick ($N_{\rm H} = 1.25 \times 10^{24}$ cm$^{-2}$, Figure \ref{comp_highE}, left) to heavier obscuration levels ($N_{\rm H} = 5 \times 10^{24}$ cm$^{-2}$, Figure \ref{comp_highE}, right). These observed differences are due to the paths transversed by the X-ray photons in the different geometries: in the toroidal model, much of the incident radiation enters the obscurer obliquely, reducing the incidence of multiple Compton scatterings since the interactions are near the surface, increasing the probability of escape. In a spherical distribution of matter, on the other hand, the X-ray photons experience the full radial distribution of matter and are forced into more Compton scatterings before escaping the medium.

\section{X-ray Spectral Fitting Description}

\subsection{Spherical Absorption Model: BNSphere}
The BNSphere model as applied in this paper is represented in XSpec as:

\begin{eqnarray}
  \mathrm{model} = & const_1 \times phabs \times (apec + {\mathrm {atable\{sphere.fits}\}} + \nonumber \\
  & const_2 \times zpow)
  \end{eqnarray}

Here, the \textsc{const$_1$} factor refers to the cross-calibration constant, which is frozen to one for the {\it Chandra} dataset and left free for the {\it NuSTAR} FPMA and FMPMB spectra. \textsc{phabs} represents the absorption from our Galaxy along the line-of-sight to NGC 4968 \citep[$N_{\rm H, Gal} = 9 \times 10^{20}$ cm$^{-2}$,][]{HI} and \textsc{apec} models the thermal emission at soft energies (below $\sim2$ keV). The {\textsc sphere.fits} file contains the table spherical absorption model of \citet{brightman}, where the fitted $N_{\rm H}$ represents the radial column density ($N_{\rm H,radial}$). The \textsc{zpow} model component represents the scattered AGN continuum that ``leaks'' through the spherical distribution of matter: to preserve the self-consistency of the model, $\Gamma$ and the power law normalization are linked to those values from the spherical model, while the \textsc{const$_2$} factor represents the fraction of leaked light. As long as this scattered fraction is small, the self-consistency of the spherical absorption model is preserved.

The spectral fits for a fixed solar abundance of Fe and with the Fe abundance as a free parameter are shown in the main text in Figure \ref{spec_fits}.

\subsection{Uniform Toroidal Absorption Model: Coupled MYTorus}
The ``default'' (coupled) MYTorus model is implemented in XSpec as:

\begin{eqnarray}
  \begin{array}{ll}
  \mathrm{model} = & const_1 \times phabs \times \nonumber \\
  & (apec + zpow_1 \times \mathrm{etable\{mytorus\_Ezero\_v00.fits\}} + \nonumber \\
  & const_2 \times \mathrm{atable\{mytorus\_scatteredH200\_v00.fits\}} + \nonumber \\
  & const_3 \times  \nonumber \\
  & \mathrm{atable\{mytl\_V000010nEp000H200\_v00.fits\}} + \nonumber \\
  & const_4 \times zpow_2)
  \end{array}
\end{eqnarray}

\textsc{const$_1$} refers to the cross-calibration normalization among the spectra, \textsc{phabs} is frozen to the Galactic absorption value, and \textsc{apec} accommodates thermal emission below 2 keV. Here \textsc{zpow$_1$} represents the intrinsic, zeroth-order AGN continuum which is attenuated by absorption in the {\textsc mytorus\_Ezero\_v00.fits} XSpec table model (\textsc{MYTorusZ}). The Compton scattered emission is modeled by the \textsc{mytorus\_scatteredH200\_v00.fits} table model (\textsc{MYTorusS}), with the fluorescent Fe K$\alpha$ and Fe K$\beta$ line emission accommodated by \textsc{mytl\_V000010nEp000H200\_v00.fits} (\textsc{MYTorusL}). 

The \textsc{MYTorusL} and \textsc{MYTorusS} table models we use have a termination energy of 200 keV, well beyond the energy coverage of the {\it NuSTAR} spectra. \textsc{const$_2$} represents the normalization of the Compton-scattered emission ($A_S$) and \textsc{const$_3$} reflects the fluorescent line emission normalization ($A_L$): $A_S \equiv A_L$ throughout, which is required to preserve the self-consistency of the model. $A_S$ captures several factors that cannot be disentangled within the model, including time delays between the direct AGN continuum and scattered emission from the torus, the covering factor of the torus, and elemental abundances different from those used in the Monte Carlo simulations that derived the MYTorus table models.

\textsc{zpow$_2$} models emission from light that escapes through the opening of the torus and scatters off a distant, optically-thin medium; $\Gamma$ and the power law normalization from this component are linked to that of \textsc{zpow$_1$}. Here, the scattering fraction ($f_{\rm scatt}$) is given by \textsc{const$_4$}. $\Gamma$ can range from 1.4 to 2.6, the limits on $N_{\rm H}$ are 10$^{22}$ to 10$^{25}$ cm$^{-2}$, and the Fe abundance is fixed to solar (i.e., it is not a free parameter).

\begin{figure*}
  \begin{center}
  \includegraphics[scale=0.2]{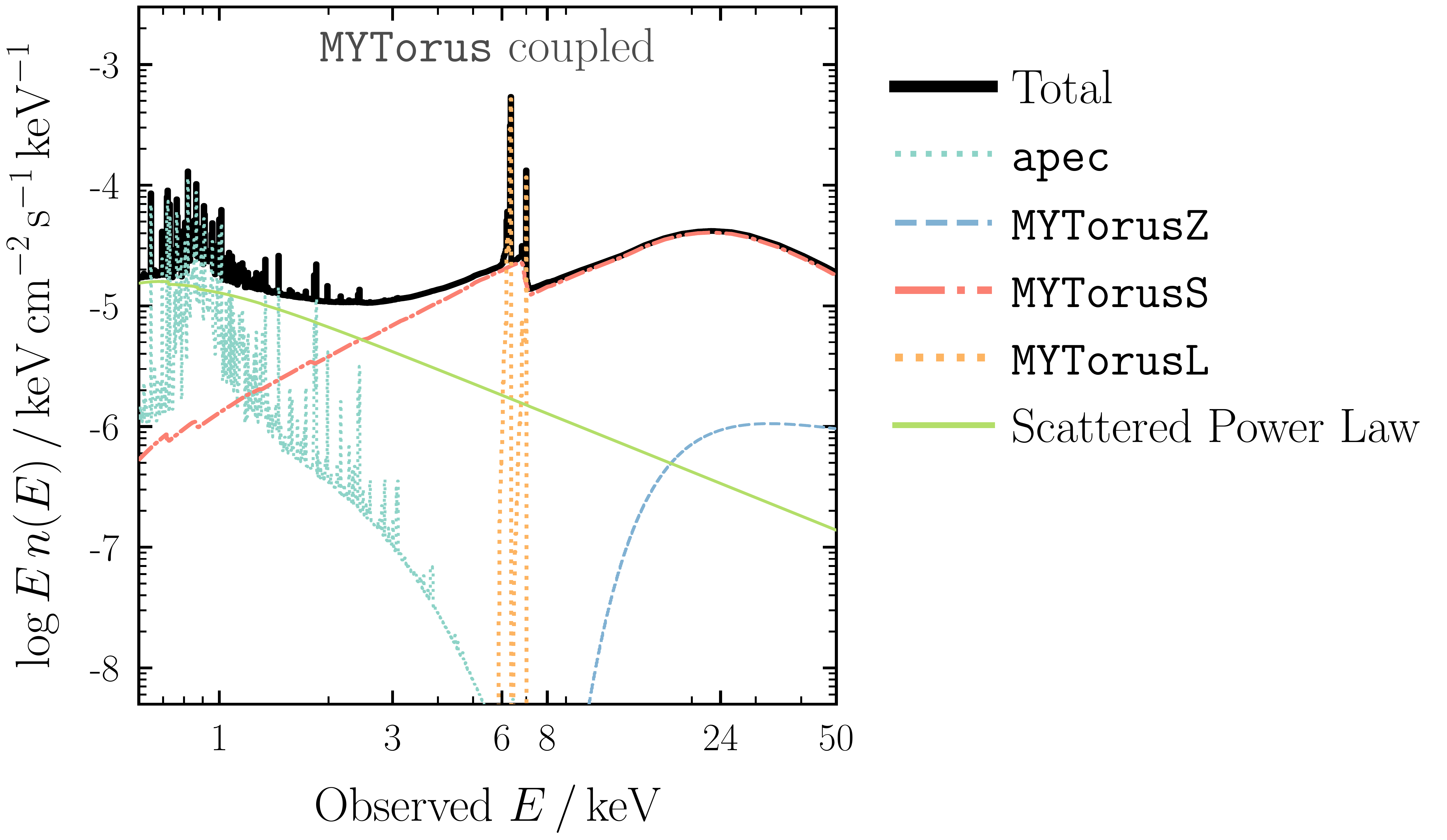}
  \caption{\label{mytor_coup_modcomp} Best fit MYTorus model is depicted by the black solid line, with the individual model components shown separately: \textsc{apec} - dotted cyan line; zeroth-order continuum (\textsc{MYTorusZ}) - blue dashed line; Compton scattered component (\textsc{MYTorusS}) - red dot-dashed line; fluorescent line emission (\textsc{MYTorusL}) - orange dotted line; scattered power law - green solid line. The spectrum is reflection dominated above 2 keV, while scattering of the intrinsic AGN continuum off a distant optically-thin medium is the primary emission below 2 keV.}
  \end{center}
\end{figure*}

In Figure \ref{spec_fits} in the main text, we show this MYTorus fit to the X-ray spectra of NGC 4968. We show the contributions of the individual model components in Figure \ref{mytor_coup_modcomp}, where the spectrum is dominated by the reprocessed AGN emission (``reflection dominated'') above 2 keV.

\subsection{``Decoupled'' MYTorus Model}
 The details for setting up the decoupled mode of the MYTorus model are described in \citet{yaqoob2012}. In brief, the column density associated with the model component \textsc{MYTorusZ} measures line-of-sight absorption ($N_{\rm H,los,mytorus}$) due to the X-ray reprocessor. The inclination angle associated with \textsc{MYTorusS} (and \textsc{MYTorusL}) is fixed to 0$^{\circ}$ to mimic backside reflection from far-side material, and the associated obscuring column density represents the global average ($N_{\rm H,global}$). As before, the normalization and photon index of the various components are linked to preserve the self-consistency of the model, so that all observed components originate from the same X-ray emitting source. The normalization of the fluorescent line emission ($A_L$) remains linked to the normalization of the Compton scattered component ($A_S$).

\begin{figure*}
\begin{center}
    \includegraphics[scale=0.2]{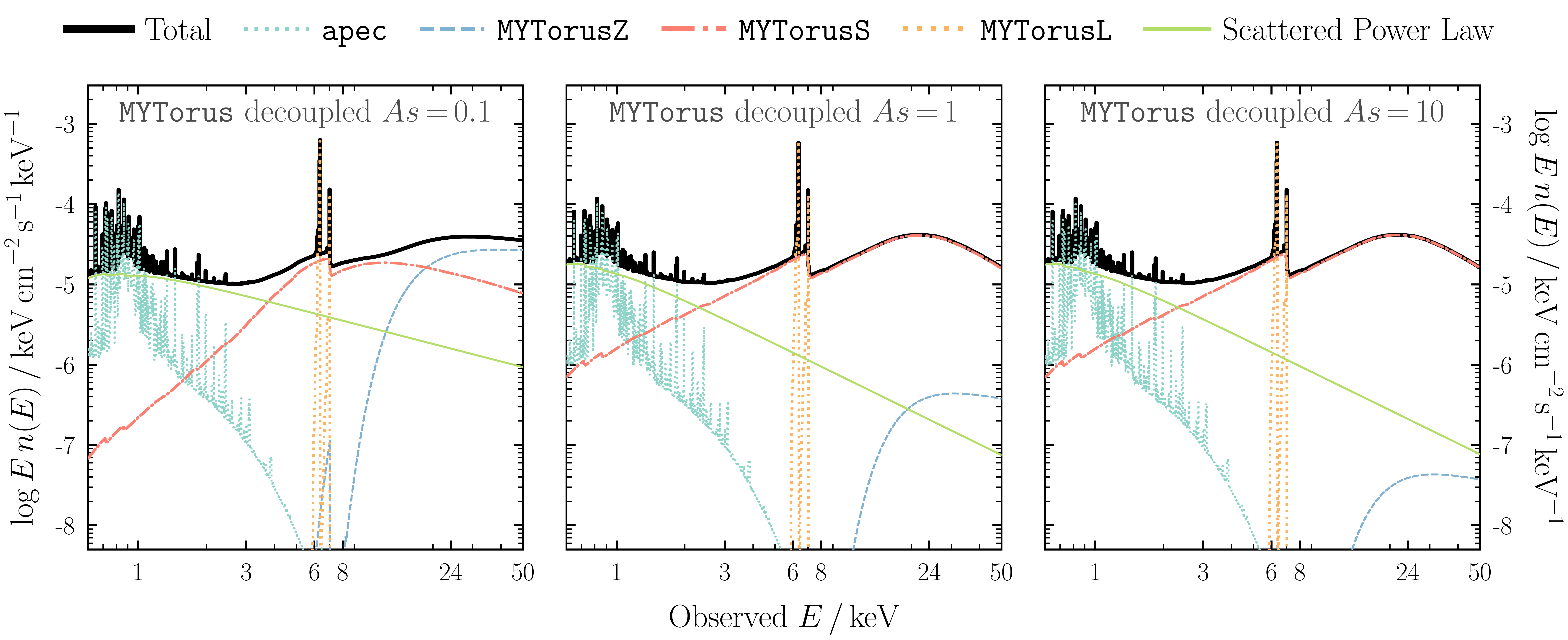}
  \caption{\label{mytor_decoup_modcomp} Individual model parameters for the decoupled MYTorus spectral fits to the {\it Chandra} and {\it NuSTAR} spectra of NGC 4968, where the line-of-sight and global column densities are fit indepenently, for ({\it left}) $A_S$ = 0.1, ({\it middle}) $A_S$ = 1, and ({\it right}) $A_S$ = 10.  Different values of the normalization for the Compton scattered emission ($A_S$) affect the relative contribution of the direct continuum (oranged dashed line) and the Compton scattered component (red dot-dashed line) to the overall emission above 10 keV. For $A_S=0.1$, the spectrum is transmission dominated above 20 keV while it is reflection dominated for higher $A_S$ values. Color and line coding is the same as Figure \ref{mytor_coup_modcomp}.}
   \end{center} 
\end{figure*}

The model fits where $A_S$ = 0.1, 1, and 10 are shown in Figure \ref{spec_fits} in the main text. As Figure \ref{mytor_decoup_modcomp} shows, the models for $A_{S}$ = 1 and $A_{S}$ = 10 indicate that the spectrum is reflection-dominated above 2 keV, with a negligible contribution from the direct component, while the transmitted component becomes the primary emission above 20 keV in the model where $A_{S}$ = 0.1 (``transmission dominated''). If the spectrum was transmission dominated at high X-ray energies, we would expect to observe X-ray variability in this regime.

\subsection{borus02 Model}
We use version 170323a of the borus02 model, where the fit parameters are the covering factor ($C_{\rm tor}$; cosine of the torus opening angle), cosine of the torus inclination angle (cos ($\theta_{\rm inc}$)), the logarithm of the torus global average column density (Log($N_{\rm H,global}$)), and parameters associated with an assumed AGN continuum of a power law with an exponential cutoff (i.e., $\Gamma$, cutoff energy, and powerlaw normalization). For reference, the allowable ranges for the borus02 model parameters are: 1.4 $< \Gamma <$ 2.6, 0.1 $ < C_{\rm tor} < $ 1  (very low covering factor to complete covering, or spherical geometry), 0.05 $<$ cos $\theta_{\rm inc} <$ 0.95 (edge-on to face-on orientation), and 22 $<$ log($N_{\rm H,global}$/cm$^{-2}$) $<$ 25.5. Note that with the borus02 model, the cosine of the inclination angle and the log of the column density are fitted, in contrast to the MYTorus model where we fit the inclination angle and column density directly.

Our borus02 model setup in XSpec is as follows:
\begin{eqnarray}
  \begin{array}{ll}
  \mathrm{model} = & const_1 \times phabs \times \nonumber \\
  & (apec + \mathrm{atable\{borus02\_v170323a.fits\}} + \nonumber \\
  & \mathrm{zphabs} \times \mathrm{cabs} \times \mathrm{cutoffpl_1} + \nonumber \\
  & const_2 \times \mathrm{cutoffpl_2}
  \end{array}
\end{eqnarray}

Similar to our modeling above, \textsc{const$_1$} accounts for the cross calibration normalization among instruments, \textsc{phabs} is frozen to the Galactic $N_{\rm H}$ value, and  \textsc{apec} models thermal emission. Since we restrict the lower energy bound on the {\it Chandra} spectrum to be above 1 keV, a significant fraction of the \textsc{apec} model component is not constrained by the data. We therefore freeze the \textsc{apec} energy (kT) and normalization to the best fit values found from the MYTorus model so that we account for the contribution of thermal emission to the spectrum at soft energies.

The borus02 table model includes the power law continuum, torus column density, inclination angle, and covering factor. Here, we freeze the cosine of the inclination angle of the torus to maximum allowed value of 0.95 to be consistent with the MYTorus decoupled model set-up where we model Compton-scattering off the backside of the torus.

\textsc{cutoffpl$_{1}$} represents the intrinsic AGN continuum attenuated by the \textsc{zphabs} and \textsc{cabs} absorption models, which account for the line-of-sight absorption and Compton scattering losses, respectively; these two column densities are linked so the line of sight attenuation from absorption and scattering is treated consistently ($N_{\rm H,los,borus}$). \textsc{cutoffpl$_{2}$} represents the scattered AGN light, where the scattering fraction is parameterized by \textsc{const$_{2}$}. Throughout, the cutoff powerlaw parameters are linked to those of the borus02 model to enforce self-consistency when modeling the AGN continuum. We freeze the cutoff energy to 200 keV but note that the X-ray spectra are not of high enough quality to measure the termination energy.

We allow the covering factor to be a free parameter in the fit, though we note that the borus02 model does not account for other effects that imprint signatures on the X-ray spectrum, such as time delays between the direct X-ray emission and scattered component and different elemental abundances than those assumed by the model. In the MYTorus model, all of these unknowns as well as the torus covering factor are encapsulated by $A_{S}$, the normalization between the direct and reprocessed emission. Within the confines of the borus02 model,  the fitted covering factor is degenerate with these other effects. Furthermore, adding extra material into the line-of-sight via the \textsc{zphabs} $\times$ \textsc{cabs} component when the torus also intercepts the direct continuum results in a geometry that is different than the one defined via the borus02 Monte Carlo simulations in which the covering factor is tabulated. This break in the self-consistency of the model impacts the measured covering factor in ways that are not quantified.

\begin{figure*}
  \begin{center}
  \includegraphics[scale=0.2]{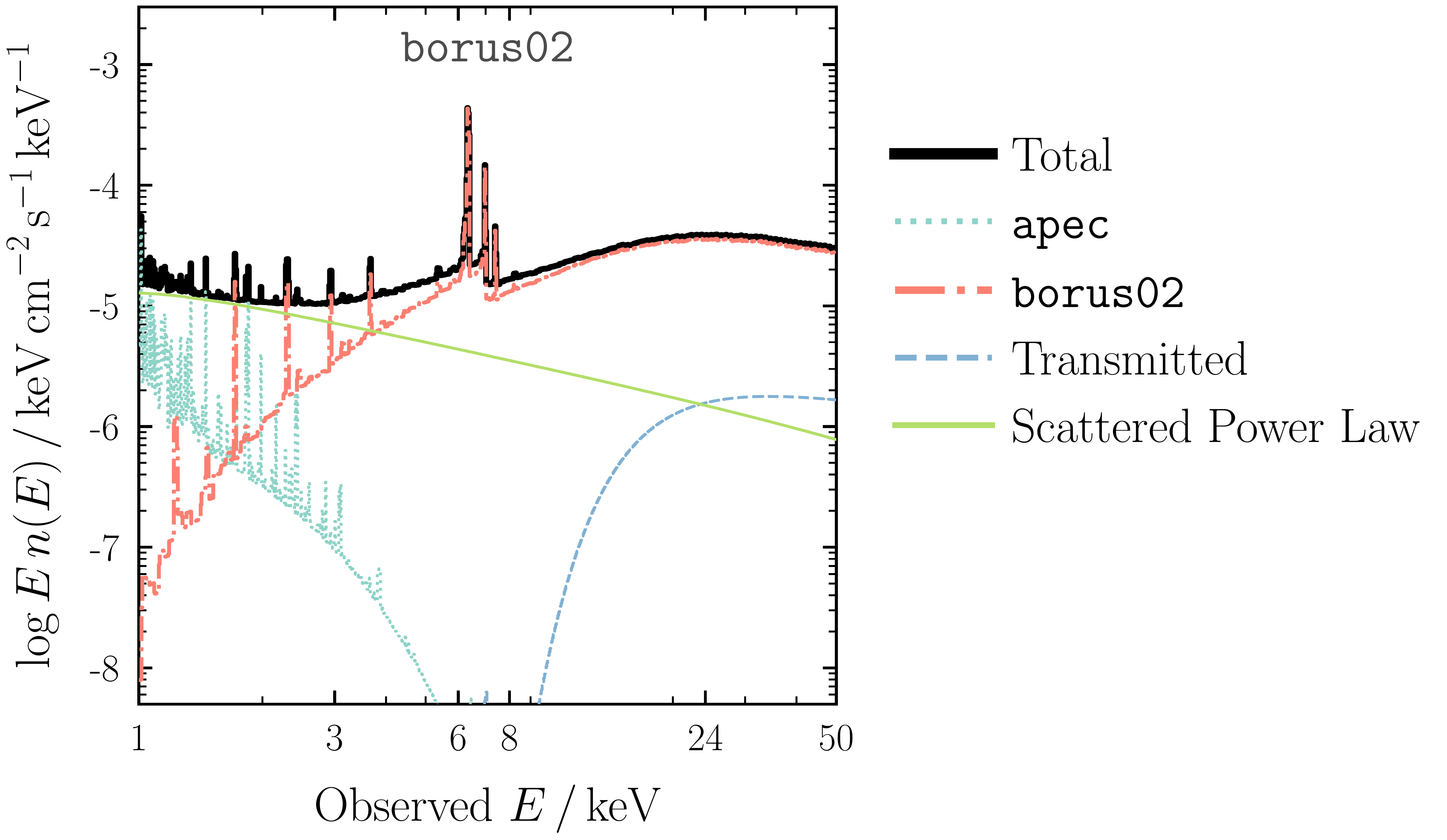}
  \caption{\label{borus_modcomp} Individual model components for the borus02 fit, where the line style and color coding is as follows: sum of model components - black line; \textsc{apec} - dotted cyan line; transmitted continuum - blue dashed line; Compton scattered component - red dot-dashed line; scattered power law - solid green line. Above 2 keV, the spectrum is dominated by the reprocessed (Compton-scattered) emission (red dot-dashed line) from the X-ray obscurer.}
  \end{center}
  \end{figure*}

The borus02 model fit to the {\it Chandra} and {\it NuSTAR} spectra of NGC 4968 is shown in Figure \ref{spec_fits} in the main text. The contributions of the individual model components to the total spectrum are shown in Figure \ref{borus_modcomp}, demonstrating that according to this model, the spectrum is reflection dominated above 2 keV.

\end{document}